\documentclass[final,hidelinks,onefignum,onetabnum]{siamart251216}

\usepackage{amssymb,mathtools}
\usepackage{booktabs,tabularx}
\usepackage{xspace}
\usepackage{tikz}
\usetikzlibrary{arrows.meta,positioning,calc}

\newsiamremark{remark}{Remark}
\newcommand{\Mix}{\mathsf{M}}
\newcommand{\Rot}{\operatorname{Rot}}
\newcommand{\rt}{\operatorname{rt}}
\newcommand{\dist}{\operatorname{dist}}
\newcommand{\diam}{\operatorname{diam}}
\newcommand{\whp}{with high probability\xspace}

\title{Makespan Bounds for Synchronous Pebble Motion on Graphs}
\author{Lucy McManamon\thanks{School of Electrical Engineering and Computer Science, Ohio University, Athens, OH, USA
  (\email{lm845822@ohio.edu}).}
  \and Jingjin Yu\thanks{Computer Science, Rutgers University, New Brunswick, NJ, USA
  (\email{jingjin.yu@cs.rutgers.edu}).}}
\headers{Makespan Bounds for Multiagent Path Finding}{L. McManamon and J. Yu}
\hypersetup{
  pdftitle={Makespan Bounds for Multiagent Path Finding on Graphs},
  pdfauthor={Lucy McManamon and Jingjin Yu}
}

\begin{document}
\maketitle
\begin{abstract}

We study the largest minimum makespan of labeled pebble motion on graphs under a model that permits synchronous pebble movement. While classical pebble motion on graphs allows only one pebble to move at a time, our model permits any number of pebbles to move to an adjacent vertex, provided that no two pebbles collide. At full occupancy every vertex is assigned a pebble and a legal round consists of disjoint cycle rotations. We prove a series of makespan bounds for this problem on graphs with varying structure and pebble density. For cycles with a chord, we extend a known quadratic bound at span two to every fixed span. For random cubic graphs, we convert permutation routing via matchings into legal schedules in two ways, applying cycle-cover batching at full occupancy and disjoint vacancy-delivery paths below full occupancy. With high probability, these yield upper bounds of $O(\log^9 n)$ at full occupancy and $O_\rho(\log^3 n)$ at every fixed pebble density $0<\rho<1$. Both cases have an $\Omega(\log n)$ lower bound. For simple 3-connected planar cubic graphs, we prove a tight linear class maximum at every pebble count. We extend the results to increasing degree, common-neighbor conditions, and vertex connectivity. The results describe how graph connectivity, cycle structure, expansion, planarity, and occupancy affect worst-case routing time.

\end{abstract}


\section{Introduction}

In pebble motion on graphs (PMG), labeled pebbles occupy distinct vertices of a finite graph, with a distinct target vertex assigned to each pebble. Classical PMG moves one pebble at a time into a vacancy \cite{KornhauserMillerSpirakis1984}. We study a synchronous variant of PMG, where in one round each pebble may wait or move along one edge, so long as no two pebbles swap across an edge or move to the same vertex. This is the underlying model for Multiagent Path Finding (MAPF), which asks for a sequence of collision-free rounds to route agents to their targets \cite{SternEtAl2019}. We study an extremal question: for a given graph and number of pebbles, how large can the minimum number of rounds be over all initial and target configurations?

At full occupancy, when the number of pebbles equals the number of vertices, this is a permutation-routing problem generated by rotations of simple cycles. A legal round consists of rotations on vertex-disjoint cycles. Below full occupancy, shifting pebbles along directed paths ending at vacancies is also a legal movement. These two types of moves connect the problem to permutation groups, classical pebble motion, and routing via matchings, but their parallel costs differ. In particular, a transposition of adjacent pebbles is a permitted operation in routing via matchings and is forbidden in one round of synchronous PMG. Our upper bounds translate a schedule of transpositions along the edges of a matching into legal rotations or vacancy-ending path shifts. We call such a translation a \emph{compiler}; its mathematical content is a bound on the number of legal rounds required for each matching layer.

Reachability alone gives little information about this parallel cost. Movement on a fully occupied cycle preserves cyclic order. Adding one chord to the cycle makes every permutation reachable but can leave a quadratic worst-case makespan. To obtain faster schedules, one must control how local rearrangements interfere. We develop three ways to do so: short cycle covers, expansion-assisted vacancy delivery, and the facial structure of planar cubic graphs.

Computing an optimal schedule is difficult even for restricted graph families \cite{YuLaValle2013,DemaineEtAl2019,FioravantesEtAl2024}. Here the objective is an extremal bound over all configuration pairs on a specified graph. Thus a result for a typical random graph must still handle every labeling of that graph.

\paragraph{Main results and methods}
Theorem~\ref{thm:spectrum} states principal makespan bounds, and Table~\ref{tab:spectrum} distinguishes their graph and occupancy quantifiers. Four arguments underpin these results.
\begin{itemize}
\item \emph{The chord-span obstruction.} We use a potential on triples of labels to extend the span-two chorded-cycle example of Foerster et~al.~\cite{FoersterEtAl2017} to every chord span $s$ (Theorem~\ref{thm:chorded-lower}). Our key estimate bounds the change of cyclic order in one round by a quantity proportional to $s$.
\item \emph{Cycle-cover transfer.} We select local rotation supports from a cycle cover and bound how many meet at a vertex (Theorem~\ref{thm:cycle-cover}). The three-rotation identity of Yu and Rus~\cite{YuRus2015} is used to perform local swaps. Our contribution is the support selection and simultaneous scheduling. Theorem~\ref{thm:barrier} gives a batching obstruction for this particular method of assigning independent supports.
\item \emph{Vacancy transfer.} On cubic expanders that are not fully occupied, we use disjoint paths to deliver two vacancies to many matching edges at once. This allows for local exchanges to be executed simultaneously. Then, vacancy delivery is reversed to restore all other pebbles. Theorem~\ref{thm:variable-hole} quantifies the cost of the transfer in terms of the total number of holes and neighborhoods that are not locally tree-like. Our contribution is a new method of parallel delivery in order to perform established swap operations.
\item \emph{Facial transfer.} For simple 3-connected planar cubic graphs, we use a simultaneous face-boundary identity and an elementary decomposition of the dual to give constant overhead per matching layer. Prism graphs give the matching linear class lower bound at every positive pebble count (Theorem~\ref{thm:planar-cubic}).
\end{itemize}

\begin{figure}[t]
\centering
\begin{tikzpicture}[
  box/.style={draw,line width=1pt,rounded corners,align=center,text width=3.45cm,inner sep=5pt,font=\footnotesize},
  arrow/.style={-{Latex},line width=1pt},node distance=8mm and 4mm]
  \node[box,text width=10.95cm] (routing) {Permutation routing via matchings: a schedule with at most $\rt(G)$ layers};
  \node[box,below=of routing] (holes) {Vacancies + expansion \\ Disjoint delivery paths\\Theorem~\ref{thm:variable-hole}};
  \node[box,left=of holes] (cycles) {Full occupancy + cycle cover\\Three-rotation identity\\Theorem~\ref{thm:cycle-cover}};
  \node[box,right=of holes] (faces) {Full occupancy + plane embedding\\Simultaneous facial identity\\Lemma~\ref{lem:facial-star}};
  \node[box,below=of cycles] (full) {Random cubic\\$O(\log^9 n)$ rounds};
  \node[box,below=of holes] (sparse) {Random cubic, fixed $\rho<1$\\$O_\rho(\log^3 n)$ rounds};
  \node[box,below=of faces] (planar) {3-connected planar cubic\\$O(n)$ rounds};
  \draw[arrow] (routing.south -| cycles.north) -- (cycles.north);
  \draw[arrow] (routing) -- (holes);
  \draw[arrow] (routing.south -| faces.north) -- (faces.north);
  \draw[arrow] (cycles) -- (full);
  \draw[arrow] (holes) -- (sparse);
  \draw[arrow] (faces) -- (planar);
\end{tikzpicture}
\caption{Dependencies of the principal upper bounds. Each branch replaces a virtual matching layer by legal rounds. The two random-graph conclusions hold with high probability.}
\label{fig:dependencies}
\end{figure}
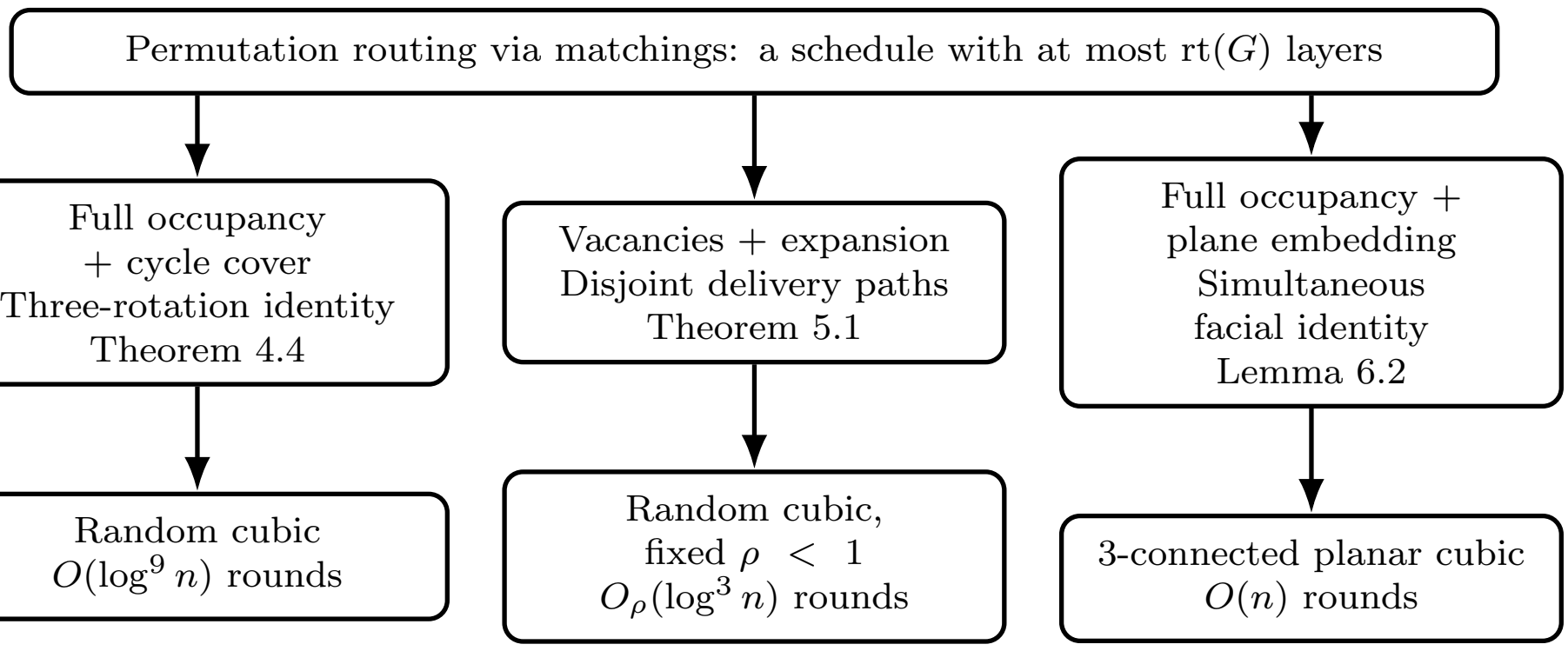

\paragraph{Related work: reachability and extremal cost}
Classical pebble motion studies reachability with vacancies \cite{KornhauserMillerSpirakis1984}, with permutation-group and parity obstructions going back to Wilson~\cite{Wilson1974}. Allowing simultaneous cycle rotations changes this reachability problem. Yu and Rus~\cite{YuRus2015} establish efficient feasibility tests and a universal quadratic group-diameter bound in the rotation model. Foerster et~al.~\cite{FoersterEtAl2017} classify solvability at full occupancy and prove a tight quadratic bound for the span-two chorded cycle. Surynek~\cite{Surynek2019} uses the name TROT for rotation variants of MAPF. Related rotation puzzles are studied by Sai et~al.~\cite{SaiVigliettaUehara2022}. Dhar et~al.~\cite{DharEtAl2026} consider universal solvability under path shifts and rotations, counting a whole shift or rotation as one atomic move. Our synchronous makespan also accounts for which operations can run in parallel.

\paragraph{Related work: routing models}
Routing via matchings permits simultaneous exchanges on a matching \cite{AlonChungGraham1994,HornPurcilly2020,Nenadov2023}. We use its bounds as inputs and calculate the extra cost of enforcing the no-swap rule. Sequential token swapping instead counts individual adjacent exchanges \cite{MiltzowEtAl2016,KawaharaSaitohYoshinaka2019}. Circular sorting permits adjacent exchanges in a cyclic order considered modulo rotation. Its exact diameter \cite{AdinAlonRoichman2025} gives a refinement of our chorded-cycle lower bound. Specifying the generators, the cost of a move, and which moves can run simultaneously is essential in each comparison.

\paragraph{Related work: vacancy-based algorithms}
Surynek's BIBOX algorithm~\cite{Surynek2009BIBOX} treats densely occupied biconnected noncycle graphs with at least two vacancies, using a loop decomposition and vacancy-assisted exchanges. It gives an $O(n^3)$ algorithmic bound and subsequently parallelizes the resulting moves. While its destination must be empty at the start of a step, our model also allows following an agent that departs in the same round. BIBOX's complexity bound therefore concerns a different cost from optimal synchronous makespan. More directly, Luna and Bekris's Push and Swap~\cite[Section~3.1, Algorithm~3]{LunaBekris2011PushSwap} clears two neighbors of a degree-three vertex, exchanges two agents, and reverses the preparation to restore other labels. We use this local idea and prove an expansion-controlled bound on how many such exchanges can be prepared and executed together.

\paragraph{Related work: simulating abstract rearrangements}
The strategy of realizing a permissive rearrangement by legal motions also has grid-specific precedents. Guo and Yu~\cite{GuoYu2024Expected} simulate row and column shuffles by legal MAPF motions, including local three-step swap gadgets. Their arbitrary-instance full-occupancy bounds and random-placement approximation guarantees have different quantifiers. Gozon and Yu~\cite{GozonYu2024Sliding} simulate row and column shuffles for generalized sliding-tile puzzles, where corner following is forbidden: a tile entering a just-vacated cell must move in the departing tile's direction. Our graph model allows such turns. These works motivate the same general translation strategy. The graph-theoretic content here is the control of support intersections, facial batching, and disjoint vacancy delivery. Section~\ref{app:lattice} records grid bounds as recovered benchmarks.

\paragraph{Organization}
After the model and principal statements in Section~\ref{sec:model}, Section~\ref{sec:lower} proves the lower bounds. Sections~\ref{sec:upper}--\ref{sec:planar} develop the three transfers displayed in Figure~\ref{fig:dependencies}. These transfers form the core of the paper. Section~\ref{sec:extensions} collects further consequences for connectivity, common neighbors, and growing-degree random regular graphs. Section~\ref{sec:discussion} separates construction costs from makespan and states remaining extremal questions.

\section{Model and Principal Results}\label{sec:model}

Let $G=(V,E)$ be a finite undirected simple graph with $n=|V|$. A configuration of $p$ labeled pebbles is an injection from the label set $[p]$ into $V$. In one legal round, each pebble waits or traverses one edge. Distinct pebbles have distinct destinations and no two pebbles may swap, or traverse the same edge in opposite directions. For an ordered pair of configurations, its \emph{minimum makespan} is the least number of legal rounds transforming the first into the second, with value $+\infty$ if the target is unreachable. Define
\[
 \Mix_p(G)=\max_{\sigma,\tau}\{\text{minimum makespan from $\sigma$ to $\tau$}\},
 \qquad \Rot(G)=\Mix_n(G).
\]
The maximum ranges over both occupied vertex sets and all placements of the labels. We set $\Mix_0(G)=0$. A graph is \emph{universally routable at full occupancy} if $\Rot(G)<\infty$.

The \emph{pebble density} is $p/n$, and the $h=n-p$ empty vertices are called \emph{holes} or \emph{vacancies}. When a fixed real density $\rho\in(0,1)$ is specified along a sequence of graphs, take $p=\lfloor\rho n\rfloor$. A statement holds \emph{with high probability} (w.h.p.) if its probability tends to one as $n\to\infty$ through the admissible values. We write $G_{n,d}$ for the uniformly random simple $d$-regular graph on $n$ labeled vertices, with $nd$ even. Constants in $O_\rho(\cdot)$ may depend on the fixed parameter $\rho$. Unsubscripted constants are independent of $n$ and of the pebble count unless another dependence is stated.

\begin{proposition}[Round decomposition]\label{prop:round-decomposition}
Relative to an initial configuration, a legal round consists exactly of a vertex-disjoint collection of directed paths with occupied nonterminal vertices and empty terminal vertices, and directed simple cycles of length at least three with occupied vertices. All other pebbles wait. In particular, at full occupancy a legal round consists of vertex-disjoint cycle rotations.
\end{proposition}
\begin{proof}
Draw an arc $u\to v$ for every moving pebble. Each vertex has indegree and outdegree at most one. If a moving pebble enters an initially occupied vertex, the resident pebble must leave, so an occupied vertex of positive indegree has outdegree one. Every nontrivial component is therefore a directed cycle or a directed path whose terminal vertex was initially empty. The no-swap rule excludes cycles of length two. Conversely, a vertex-disjoint collection of these paths and cycles defines a legal round, with all other pebbles waiting. At full occupancy there is no empty terminal vertex, so no path component occurs.
\end{proof}

A path shift is one round regardless of its length: each pebble on the path moves by one edge. The hole moves from the terminal vertex to the initial vertex of the path. This convention is essential to the vacancy transfer in Section~\ref{sec:vacancy}.

\begin{proposition}[Density and edge monotonicity]\label{prop:monotonicity}
For every graph $G$ and $0\le p_1\le p_2\le n$,
\[
 \Mix_{p_1}(G)\le\Mix_{p_2}(G).
\]
In particular, $\Mix_p(G)\le\Rot(G)$ for every $p\le n$. If $H$ is a spanning subgraph of $G$, then $\Mix_p(G)\le\Mix_p(H)$ for every $p$.
\end{proposition}
\begin{proof}
If $\Mix_{p_2}(G)=+\infty$, the density inequality is immediate. Otherwise extend any $p_1$-pebble instance by adding $p_2-p_1$ distinctly labeled dummy pebbles at arbitrary initial holes and assigning them to distinct target holes. Follow a schedule for the enlarged instance and erase the dummy pebbles throughout. The remaining pebbles still have distinct destinations, and erasing moves cannot introduce an opposite-edge exchange. Thus every projected round is legal. Equivalently, the paths and cycles of Proposition~\ref{prop:round-decomposition} break into legal paths when dummy pebbles are removed. The projected schedule solves the original instance within $\Mix_{p_2}(G)$ rounds. Finally, every round legal in a spanning subgraph $H$ is legal in $G$, which proves edge monotonicity.
\end{proof}

For comparison, the \emph{matching-routing number} $\rt(G)$ is the largest, over all permutations of labels occupying $V$, of the minimum number of rounds when a round may exchange the endpoint labels on every edge of a matching. These exchanges are virtual operations in our proofs, as the no-swap rule prohibits a single exchange across an edge. The transfers below replace a matching round by a bounded number of legal rounds.

\begin{theorem}[Principal makespan bounds]\label{thm:spectrum}
The following statements hold.
\begin{enumerate}
\item If $G$ is a connected 2-regular graph, then $\Rot(G)=+\infty$. There are also 2-edge-connected graphs with $\Rot(G)=+\infty$.
\item Let $H_{n,s}$ be $C_n$ with one chord whose endpoints have cyclic distance $s$, where $2\le s\le\lfloor n/2\rfloor$. Uniformly over these spans,
\[
 \frac{n(n-1)}{6s}\le\Rot(H_{n,s})=O(n^2).
\]
For every fixed $s\ge2$, this gives $\Rot(H_{n,s})=\Theta(n^2)$. Moreover,
\[
 \max_{\substack{|V(G)|=n\\G\text{ 2-vertex-connected, noncycle}}}\Rot(G)=\Theta(n^2).
\]
\item Connected cubic graphs can have $\Rot(G)=+\infty$. Every connected bridgeless cubic graph is universally routable at full occupancy and satisfies $\Rot(G)=O(n^2)$.
\item For even $n$, with high probability,
\[
 \Omega(\log n)\le\Rot(G_{n,3})\le O(\log^9 n).
\]
For every fixed $0<\rho<1$, with high probability,
\[
 \Omega(\log n)\le\Mix_{\lfloor\rho n\rfloor}(G_{n,3})\le O_\rho(\log^3 n).
\]
\item For every even $n\ge4$ and every $1\le r\le n$,
\[
 \max_{\substack{|V(G)|=n\\G\text{ simple, 3-connected,}\\\text{planar, and cubic}}}\Mix_p(G)=\Theta(n),
\]
with constants independent of $p$.
\end{enumerate}
Every full-occupancy upper bound also bounds $\Mix_p(G)$ for all $p\le n$.
\end{theorem}

\begin{table}[t]
\caption{Principal bounds and their quantifiers. A class maximum is an upper bound for every graph in the class and a matching lower bound for a family of witnesses. The random cubic rows state established bounds, not known optimal ones.}
\label{tab:spectrum}
\centering
\small
\begin{tabular}{@{}>{\raggedright\arraybackslash}p{0.28\textwidth}>{\raggedright\arraybackslash}p{0.18\textwidth}>{\raggedright\arraybackslash}p{0.23\textwidth}>{\raggedright\arraybackslash}p{0.19\textwidth}@{}}
\toprule
Graph family & Occupancy & Makespan bound & Scope \\
\midrule
Cycles & full & $+\infty$ & every graph \\
One-chord cycles, fixed span & full & $\Theta(n^2)$ & each fixed span \\
Connected cubic with a bridge & full & $+\infty$ & every such graph \\
Connected bridgeless cubic & full & $O(n^2)$ & upper bound \\
Random cubic & full & $\Omega(\log n)$ to $O(\log^9 n)$ & w.h.p.; gap open \\
Random cubic & $p=\lfloor\rho n\rfloor$, fixed $0<\rho<1$ & $\Omega(\log n)$ to $O_\rho(\log^3 n)$ & w.h.p.; gap open \\
Simple 3-connected planar cubic & every $1\le p\le n$ & $\Theta(n)$ & class maximum \\
\bottomrule
\end{tabular}

\end{table}

\paragraph{Proof dependencies and scope}
The reachability obstructions in items (1) and (3), and the chord-span lower bound in item (2), are proved in Section~\ref{sec:lower}. The quadratic upper bound for 2-vertex-connected noncycles is due to Yu and Rus~\cite{YuRus2015}. In lemma~\ref{lem:bridgeless-2conn}, we apply it to bridgeless cubic graphs. The span-two quadratic lower bound is due to Foerster et~al.~\cite{FoersterEtAl2017}. Item (4) combines the counting lower bound of Proposition~\ref{prop:counting} with Corollaries~\ref{cor:random-cubic} and~\ref{cor:fixed-density}. Item (5) is Theorem~\ref{thm:planar-cubic}.

The planar statement is a class maximum. The cubic statements distinguish bridges, expansion, and planarity. For random cubic graphs the logarithmic lower bound leaves substantial gaps at both densities, and the two upper bounds do not establish an asymptotic separation between optimal makespans.

Section~\ref{sec:extensions} collects bounds in terms of vertex connectivity and common neighbors, together with growing-degree random-regular consequences. In particular, three common neighbors per vertex pair suffice for the codegree theorem. These are consequences of the same local transposition approach.

\section{Lower Bounds}\label{sec:lower}

\subsection{Generic Obstructions}
We begin with observations on three generic obstructions: distance, cut throughput, and state-space counting. Throughout, $\pi$ denotes an instance with initial vertex $s_x$ and target vertex $t_x$ for each pebble $x$, and $T(\pi)$ its minimum makespan.

\begin{proposition}[Distance bound]\label{prop:dist}
$T(\pi)\ge\max_x \dist_G(s_x,t_x)$; consequently, $\Mix_p(G)\ge\diam(G)$ for $1\le p\le n$, where $diam(G) = \infty$ if $G$ is disconnected.
\end{proposition}

\begin{proof}
For the first claim, a pebble traverses at most one edge per round. For the second claim, let $(u, v)$ be a diametral pair of vertices. Construct an instance with $s_1 = u$ and $t_1 = v$. Assign the remaining $p-1$ labels arbitrarily.
\end{proof}

\begin{proposition}[Cut bound]\label{prop:cut}
For $S\subseteq V$, let $\delta(S)$ be its edge boundary and let $x_\pi(S)$ count pebbles initially in $S$ with targets outside $S$. If $|\delta(S)|>0$, then $T(\pi)\ge x_\pi(S)/|\delta(S)|$. At full occupancy, $T(\pi)\ge 2x_\pi(S)/|\delta(S)|$.
\end{proposition}

\begin{proof}
In one round each boundary edge can transfer at most one pebble from $S$ to $G-S$. In a rotation round, every directed cycle crosses the cut $\delta(S)$ the same number of times inward as outward. Therefore, transferring a pebble out of $S$ requires two boundary-edge traversals in total, one outward and one inward, meaning at most $|\delta(S)|/2$ pebbles leave $S$ per round.
\end{proof}

\begin{proposition}[Counting bound]\label{prop:counting}
Let $G$ have maximum degree $\Delta\ge1$, and let $1\le r\le n$. Then
\[
    \Mix_r(G)\ge
    \frac{\log (n)_r}{r\log(\Delta+1)},
    \qquad (n)_r=\frac{n!}{(n-r)!},
\]
which is $\Omega(\log n)$ for every fixed-degree family, uniformly over $1\le r\le n$.
\end{proposition}

\begin{proof}
If some placement pair is unreachable, the bound is immediate. Otherwise, from a fixed configuration each pebble has at most $\Delta+1$ options per round. Padding shorter schedules with waits shows that at most $(\Delta+1)^{rT}$ configurations are reachable within $T$ rounds. Taking $T=\Mix_r(G)$, all $(n)_p$ configurations must be reachable, giving the display. The geometric mean of the $p$ largest elements of $\{1,\ldots,n\}$ is at least the geometric mean of all $n$, so $p^{-1}\log(n)_r\ge n^{-1}\log n!=\log n-O(1)$, uniformly in $p$.
\end{proof}

\subsection{Reachability Before Makespan}

A connected 2-regular graph on n vertices is a cycle, $C_n$. At full occupancy its only nontrivial rounds are rotations of the entire cycle, so the reachable permutations form the cyclic group generated by one rotation and only $n$ of the $n!$ permutations are reachable. Therefore, $\Rot(C_n)=+\infty$.

Bridges can obstruct connected cubic graphs. Because a bridge lies on no cycle, at maximum density no rotation is able to move a pebble across it, so every target permutation sending a pebble across a bridge is unreachable. To construct a witness, take two copies of $K_4$ minus an edge, with degree-2 vertices $a_1,b_1$ in copy $1$ and $a_2,b_2$ in copy $2$. Add adjacent vertices $u,v$ with edges $ua_1,ub_1$ and $va_2,vb_2$. The resulting graph is cubic and $uv$ is a bridge, so $\Rot(G)=+\infty$, proving the first claim of Theorem~\ref{thm:spectrum}(3).

Even 2-edge-connectivity is insufficient. In a figure-eight made of two triangles sharing one vertex, every generating rotation is a 3-cycle and is thus an even permutation. Every odd target permutation is unreachable, a parity obstruction following Wilson~\cite{Wilson1974}. By contrast, full-occupancy reachability is satisfied by graphs that are 2-vertex-connected and not cycles \cite{YuRus2015}. This condition is sufficient but not necessary for individual graphs.

\subsection{A Sharp Quadratic Family}

Foerster et~al.~\cite{FoersterEtAl2017} used a potential argument to prove a quadratic lower bound on a span two chorded cycle. We retain their approach and make the bound explicit for any arbitrary chord span. A circular-sorting theorem gives a further numerical refinement at the end of this subsection.

\begin{figure}[t]
\centering
\begin{tikzpicture}[scale=0.8, every node/.style={circle,draw,minimum size=5.5mm,inner sep=1pt,font=\small}, >=Latex]
  \node (u) at (0,0) {$v_0$};
  \node (a) at (1.15,1.05) {$v_1$};
  \node (b) at (2.35,1.25) [draw=none] {$\cdots$};
  \node (v) at (3.8,0) {$v_s$};
  \node (c) at (2.35,-1.25) [draw=none] {$\cdots$};
  \node (d) at (1.15,-1.05) {$v_{n-1}$};
  \draw[line width=1pt] (u)--(a)--(b)--(v);
  \draw[line width=1pt] (u)--(d)--(c)--(v);
  \draw[very thick] (u)--(v);
  \node[draw=none,rectangle] at (1.9,0.32) {chord};
  \node[draw=none,rectangle] at (1.9,1.8) {$s$-edge arc};
  \node[draw=none,rectangle] at (1.9,-1.8) {complementary arc};
\end{tikzpicture}
\caption{The graph $H_{n,s}$ has exactly three simple cycles. All three meet at $v_0$ and $v_s$, so a synchronous round contains at most one rotation.}
\label{fig:chorded}
\end{figure}
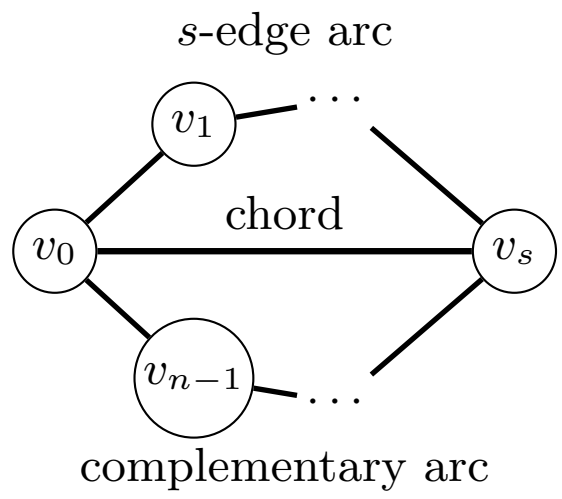

For $n\ge4$ and $2\le s\le\lfloor n/2\rfloor$, let $H_{n,s}$ be the cycle $C_n=(v_0v_1\cdots v_{n-1}v_0)$ plus the chord $v_0v_s$ (Figure~\ref{fig:chorded}). It has exactly three cycles: the outer cycle and the two cycles formed by the chord. All cycles contain both $v_0$ and $v_s$, so at most one cycle can rotate in a synchronous round.

Read the pebble labels clockwise around the outer cycle $C_n$, treating cyclic shifts as equivalent. The \emph{triple orientation} of a set of three labels $\{a,b,c\}$ is which of the two circular orders it induces, either clockwise or counterclockwise. Reversing the order of the outer cycle reverses all $\binom n3$ triple orientations. The potential argument rests on two lemmas.

\begin{lemma}[Adjacent circular interchange]\label{lem:adjacent-triples}
Interchanging two adjacent labels in a cyclic order reverses exactly $n-2$ triple orientations.
\end{lemma}

\begin{proof}
Let the adjacent labels be $a,b$. Any three label subset not containing both $a$ and $b$ induces the same circular order after the interchange. For a three label subset containing $a$, $b$, and a third label $c$, the relative circular order of $a,b,c$ is reversed. There are exactly $n-2$ choices of $c$.
\end{proof}

\begin{lemma}[Effect of one legal rotation]\label{lem:rotation-step}
Modulo cyclic shifts, one legal rotation on $H_{n,s}$ is a product of at most $s$ adjacent circular interchanges. Consequently it changes at most $s(n-2)$ triple orientations of the outer cyclic order.
\end{lemma}

\begin{proof}
Write the current outer cyclic word as $(a_0,a_1,\ldots,a_{n-1})$, indexed so that $a_0,\ldots,a_s$ occupy $v_0,\ldots,v_s$. A rotation of the outer cycle cyclically shifts the word and changes no triple orientation. A rotation of the short cycle $(v_0,v_1,\ldots,v_s)$ transforms $(a_0,\ldots,a_s)$ into $(a_s,a_0,\ldots,a_{s-1})$ or $(a_1,\ldots,a_s,a_0)$, according to orientation. Either transformation is a product of $s$ adjacent circular interchanges. A rotation of the complementary cycle maps
\begin{align*}
&(a_0,\ldots,a_{s-1},a_s,\ldots,a_{n-1})\\
&\qquad\mapsto
(a_{n-1},a_1,\ldots,a_{s-1},a_0,a_s,\ldots,a_{n-2});
\end{align*}
reading both circular words from $a_{n-1}$ shows that this moves $a_0$ across only $s-1$ adjacent labels, and the reverse orientation is symmetric. Lemma~\ref{lem:adjacent-triples} bounds the number of changed triple orientations by $s(n-2)$ in every case.
\end{proof}

\begin{theorem}[Chorded-cycle lower bound]\label{thm:chorded-lower}
For $n\ge4$ and $2\le s\le\lfloor n/2\rfloor$,
\[
    \Rot(H_{n,s})\ge \frac{\binom n3}{s(n-2)}=\frac{n(n-1)}{6s}.
\]
In particular, $\Rot(H_{n,s})=\Theta(n^2)$ for every fixed $s\ge2$, and $\Rot(H_{n,2})\ge n(n-1)/12$.
\end{theorem}

\begin{proof}
Start from the outer cyclic order $(1,2,\ldots,n)$ and choose a target configuration whose outer cyclic order is the reversal $(1,n,n-1,\ldots,2)$. All $\binom n3$ triple orientations differ between the two orders. The target is reachable since $H_{n,s}$ is 2-vertex-connected and not a cycle \cite{YuRus2015}. Since all three simple cycles share the chord endpoints, at most one rotates per round. By Lemma~\ref{lem:rotation-step} each round changes at most $s(n-2)$ triple orientations. Thus any schedule of $T$ rounds satisfies $s(n-2)\,T\ge\binom n3$. The matching $O(n^2)$ upper bound for fixed $s$ is given by the group-diameter theorem of Yu and Rus~\cite{YuRus2015}.
\end{proof}

\begin{corollary}[Sharpness of the quadratic upper bound]\label{cor:quadratic-sharp}
Let $F_2(n)$ be the largest value of $\Rot(G)$ over all 2-vertex-connected noncycle graphs on $n$ vertices. $F_2(n)=\Theta(n^2)$.
\end{corollary}

\begin{proof}
The upper bound is given by Yu and Rus~\cite{YuRus2015}. The lower bound is given by Theorem~\ref{thm:chorded-lower} with $H_{n,2}$.
\end{proof}

Thus the quadratic upper bound is tight for the extremal function $F_2$, even when vertex-disjoint rotations may execute in parallel. The span-dependent bound above generalizes the known span-two case. It does not claim a new quadratic extremal order.

\begin{remark}[Refinement by circular sorting]\label{rem:circular-sorting}
Adin, Alon, and Roichman~\cite[Theorem~1.1]{AdinAlonRoichman2025} prove that the diameter of cyclic orders under adjacent circular interchanges is $\lfloor(n-1)^2/4\rfloor$. Here an order is considered modulo rotation, as above. Choose two orders attaining that distance. By Lemma~\ref{lem:rotation-step}, any $T$-round schedule between configurations with these orders yields at most $sT$ interchanges, so
\[
 \Rot(H_{n,s})\ge \frac{\lfloor(n-1)^2/4\rfloor}{s}.
\]
Theorem~\ref{thm:chorded-lower} gives an independent bound with the same dependence on $s$. Neither estimate determines the order of $\Rot(H_{n,s})$ for all growing spans.
\end{remark}

\section{Rotation Routing at Full Occupancy}\label{sec:upper}

We start by adopting the $O(n^2)$ group diameter for 2-vertex-connected noncycles \cite{YuRus2015}. We show why it applies to all bridgeless cubic graphs, completing item (3) of Theorem~\ref{thm:spectrum}.

\begin{lemma}[Bridgeless implies 2-connected]\label{lem:bridgeless-2conn}
Every connected bridgeless cubic graph is 2-vertex-connected and is not a cycle, therefore it is universally routable with $\Rot(G)=O(n^2)$ \cite{YuRus2015}.
\end{lemma}

\begin{proof}
If a cubic graph had a cut vertex $x$, the three edges at $x$ would be distributed among at least two components of $G-x$. Some component would receive exactly one of them, and that edge would be a bridge. A connected bridgeless cubic graph therefore has no cut vertex. Finally, being 3-regular, it is not a cycle. Yu and Rus~\cite{YuRus2015} applies.
\end{proof}

The sharper upper bounds use the matching-routing number $\rt(G)$ defined in Section~\ref{sec:model}. A matching-routing round exchanges labels across edges of a matching, so its transpositions must be simulated by legal rounds. At full occupancy we simulate a transposition with three rotations on a theta graph supporting the edge. To execute many of these simulations simultaneously, we control both the size of each supporting subgraph and the number of supports containing any one vertex. We develop a different simulation for graphs with two or more vacancies in Section~\ref{sec:vacancy}.

\subsection{The first edge theta identity}

\begin{figure}[t]
\centering
\begin{tikzpicture}[scale=0.8, every node/.style={circle,draw,minimum size=5.5mm,inner sep=1pt,font=\small}, >=Latex]
  \node (u) at (0,0) {$u$};
  \node (b1) at (1.15,0) {$b_1$};
  \node (bt) at (3.15,0) {$b_t$};
  \node (w) at (4.35,0) {$w$};
  \node (p1) at (1.0,1.3) {$p_1$};
  \node (pr) at (3.3,1.3) {$p_r$};
  \node (q1) at (1.0,-1.3) {$q_1$};
  \node (qs) at (3.3,-1.3) {$q_s$};
  \draw[line width=1pt] (u)--(b1)--(bt)--(w);
  \draw[line width=1pt] (u)--(p1)--(pr)--(w);
  \draw[line width=1pt] (u)--(q1)--(qs)--(w);
  \node[draw=none,rectangle] at (2.2,0.35) {$B$};
  \node[draw=none,rectangle] at (2.2,1.72) {$P$};
  \node[draw=none,rectangle] at (2.2,-1.72) {$Q$};
\end{tikzpicture}
\caption{A first-edge theta. The transposed edge $ub_1$ need only be the first edge of branch $B$, not an entire branch.}
\label{fig:theta}
\end{figure}
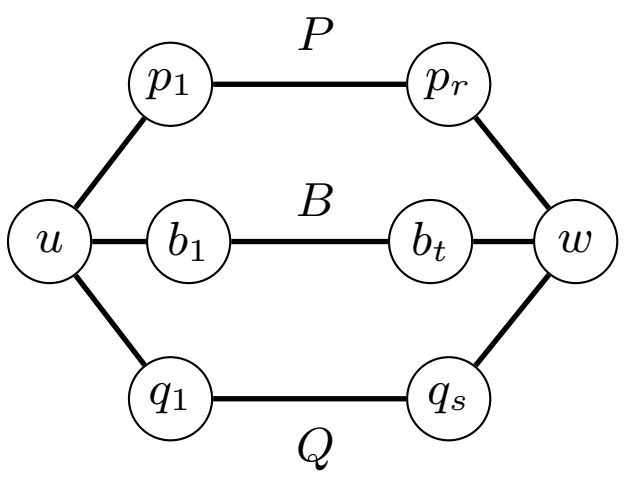

A \emph{theta graph} is the union of three internally vertex-disjoint paths, or \emph{branches}, with common endpoints, or \emph{terminals}. The chorded cycle $H_{n,s}$ in Section~\ref{sec:lower} is one example, and its chord is a branch that is a single edge. We use the three rotation exchange from the proof of Theorem~2 in Yu and Rus~\cite{YuRus2015}. The exchanged edge is the \emph{first} edge of one branch. That branch need not consist of a single edge (Figure~\ref{fig:theta}). We restate the identity and its proof to specify the supports used by the transfer theorems.

\begin{lemma}[First-edge theta identity, Yu--Rus~\cite{YuRus2015}]\label{lem:theta}
Let a theta graph have terminals $u,w$ and three internally vertex-disjoint branches
\[
 B=(u,b_1,\ldots,b_t,w),\qquad
 P=(u,p_1,\ldots,p_r,w),\qquad
 Q=(u,q_1,\ldots,q_s,w).
\]
A branch may be the single edge $uw$ (for $B$, set $b_1=w$, similarly for $P,Q$). The three cycles of the theta are defined as follows. Let $C_P$ traverse $P$ from $u$ to $w$ and $B$ backward, let $C_Q$ traverse $Q$ from $u$ to $w$ and $B$ backward, and let $C_0$ traverse $P$ forward and $Q$ backward. Writing $R_C$ for a cyclic rotation and applying the rightmost operation first,
\[
    R_{C_Q}^{-1}\,R_{C_0}^{-1}\,R_{C_P}=(u\ b_1),
\]
and every other vertex of the theta remains fixed. Thus the labels across the edge $ub_1$ are exchanged in three rotation rounds.
\end{lemma}

\begin{proof}
Track the labels of interest. Under $R_{C_P}$, the label at $u$ moves to $p_1$ and the label at $b_1$ moves to $u$. Under $R_{C_0}^{-1}$, the label at $p_1$ moves to $u$ and the label at $u$ moves to $q_1$. Under $R_{C_Q}^{-1}$, the label at $u$ moves to $b_1$ and the label at $q_1$ moves to $u$. Hence the labels of $u$ and $b_1$ are exchanged. The two relevant rotations cancel on each interior vertex of $P$ and $Q$, and on every interior vertex of $B$ beyond $b_1$. If $B$ has an interior vertex, the label at $w$ moves first to $b_t$ and finally back to $w$. If $B=uw$, then $w=b_1$ is one of the exchanged vertices. The same tracking includes $P=uw$ or $Q=uw$.
\end{proof}

\begin{theorem}[Local support compiler]\label{thm:compiler}
Choose an orientation for every edge of $G$. Suppose each oriented edge $e$ has a vertex support $H_e$ of size at most $L$ and at most $h$ legal rotation rounds supported on $H_e$ transpose its endpoint labels and fix all other labels. Suppose every vertex of $G$ lies in at most $\mu_H$ supports, counted with their edge indices even if some supports coincide. Then
\[
    \Rot(G)\le h\bigl(L(\mu_H-1)+1\bigr)\rt(G).
\]
A first-edge theta family has $h=3$.
\end{theorem}

\begin{proof}
Form the conflict graph for a matching $M$ of a matching-routing schedule, adding an edge between vertices $e$ and $f$ when their supports $H_e$ and $H_f$ are vertex-disjoint. Each support has at most $L$ vertices, and each vertex belongs to at most $\mu_H-1$ other supports, so the conflict graph has maximum degree at most $L(\mu_H-1)$, and greedy coloring uses at most $L(\mu_H-1)+1$ colors. Within one color class the rotation rounds run simultaneously in $h$ rounds. Compile all $\rt(G)$ layers.
\end{proof}

\subsection{Cubic graphs and cycle covers}

An \emph{$(\ell,c)$-cycle cover} is a family of simple cycles in which every edge appears at least once and at most $c$ times and every cycle has length at most $\ell$.

\begin{theorem}[Cycle-cover transfer]\label{thm:cycle-cover}
If a bridgeless cubic graph $G$ has an $(\ell,c)$-cycle cover, then
\[
    \Rot(G)=O(c\ell^2)\,\rt(G).
\]
\end{theorem}

\begin{proof}
We first construct one theta $H_u$ at each vertex $u$, then bound how often these thetas overlap. Let $x$, $y$, and $z$ be the neighbors of $u$. A cover cycle through a vertex uses exactly two of its incident edges. To cover all three edges incident to $u$, two cover cycles $C_u$ and $D_u$ are needed. Both covers share a common edge $uz$, with one covering $ux$ and another covering $uy$. In $G \setminus \{u\}$, the remnants of the cycles are paths that both contain $z$. Their union is connected and contains all three neighbors of $u$.

Take a spanning tree of this union and prune leaves other than the three neighbors. The resulting tree has a unique median $w_u$ at which the paths from the three neighbors meet. These paths share only $w_u$. One of $u$'s neighbors may be equal to $w_u$, and in this case a path has length 0. Adding $u$ and its three incident edges gives a theta $H_u$ with terminals $u,w_u$. It has at most $2\ell$ vertices, and every edge incident to $u$ begins one of its branches. Lemma~\ref{lem:theta} thus supplies a three-round exchange for each such edge.

To bound the load on each edge, fix a vertex $v$. At most $3c$ cover cycles contain $v$, because each uses an edge incident to $v$ and each of the three incident edges is in at most $c$ cycles. Each cover cycle can be selected as $C_u$ or $D_u$ at no more than $\ell$ roots $u$ on that cycle. Therefore, $v$ belongs to at most $3c\ell$ rooted thetas $H_u$. Orient each edge arbitrarily and consider the theta rooted at its tail. Each rooted theta is assigned to at most three edges. Consequently the edge-indexed support load is at most $9c\ell$. Apply Theorem~\ref{thm:compiler} with $h=3$, $L\le2\ell$, and $\mu_H\le9c\ell$.
\end{proof}

The random-graph consequence uses two imported bounds. Let $\lambda$ be the largest nontrivial adjacency eigenvalue of a $d$-regular graph, where $\lambda=\max_{i\ge2}|\lambda_i|$, where $d=\lambda_1,\lambda_2,\ldots,\lambda_n$ are the adjacency eigenvalues. Alon, Chung, and Graham~\cite[Theorem~7]{AlonChungGraham1994} prove
\[
    \rt(G)=O\!\left(\frac{d^2}{(d-\lambda)^2}\log^2 n\right)
    \qquad(\lambda<d).
\]
Parter and Yogev~\cite[Theorem~1; expanded version, Section~3.2]{ParterYogev2019} give every connected bridgeless graph of diameter $D$ a cycle cover with
\[
    \ell=O(D\log n),\qquad c=O(\log^3 n).
\]

\begin{corollary}[Random cubic graphs]\label{cor:random-cubic}
For even $n$ and a uniformly random simple cubic graph, \whp,
\[
    \Rot(G_{n,3})=O(\log^9 n).
\]
Consequently, by Proposition~\ref{prop:monotonicity}, \whp $\Mix_r(G_{n,3})=O(\log^9 n)$ for every $0\le r\le n$.
\end{corollary}

\begin{proof}
With high probability $G_{n,3}$ is 3-connected \cite{Wormald1981,Wormald1999}, and therefore bridgeless, has diameter $D=O(\log n)$ \cite{BollobasFernandezDeLaVega1982}, and satisfies $\lambda\le2\sqrt{2}+\varepsilon<3$ for any fixed sufficiently small $\varepsilon>0$ \cite{Friedman2008}. The two imported bounds give $\rt(G)=O(\log^2 n)$, $\ell=O(\log^2 n)$, and $c=O(\log^3 n)$. Theorem~\ref{thm:cycle-cover} gives
\begin{align*}
    \Rot(G_{n,3})
    &=O(c\ell^2)\,\rt(G)\\
    &=O(\log^3 n\cdot\log^4 n\cdot\log^2 n)
    =O(\log^9 n).
\end{align*}
\end{proof}

\paragraph{Constructing the schedule.}
Given the cover as cyclic vertex lists, the support construction uses tree searches in unions of two cycles, each with $O(\ell)$ vertices. The total preprocessing time is $O(|E(G)|c+n\ell)$: first index the cycles at their edges and vertices, then construct all rooted thetas. Given also a virtual schedule, the conflict graph for each matching layer is obtained by listing the supports incident to each vertex. Greedy coloring and the displayed rotation words complete the conversion in polynomial time. The cycle-cover algorithm of Parter and Yogev is explicit. Its distributed implementation uses $\widetilde O(n)$ communication rounds \cite[expanded version, Lemma~5]{ParterYogev2019}. Communication rounds are a separate measure from centralized running time and from MAPF makespan. The random-walk construction in the proof of Alon et~al.'s Theorem~7 likewise supplies the virtual schedule. Its sampled walk intersections can be checked before compilation.

Proposition~\ref{prop:counting} supplies the $\Omega(\log n)$ lower bound in Theorem~\ref{thm:spectrum}(4). This leaves a gap in the exponent. The next result rules out constant overhead for one particular way of compiling a matching layer. It does not determine the optimal routing exponent.

\subsection{A limitation of disjoint theta supports}

\begin{theorem}[Logarithmic batching barrier]\label{thm:barrier}
There is a constant $\gamma>0$ such that \whp a random cubic graph contains a matching $M$ of $\Theta(n)$ edges such that every first-edge theta support of every $e\in M$ has at least $\gamma\log n$ vertices. Consequently, any compiler that assigns one such support to each edge of $M$ and executes only vertex-disjoint supports in one round needs $\Omega(\log n)$ rounds for this single matching layer.
\end{theorem}

\begin{proof}
Let $g=\lfloor\frac14\log_2 n\rfloor$. We first show that \whp only $o(n)$ edges lie on cycles of length at most $g$. For random cubic graphs, the number $X_k$ of $k$-cycles and $3\le k\le g$ is
\[
 \mathbb E[X_k]
 =\frac{(n)_k6^k}{2k\prod_{j=0}^{k-1}(3n-2j-1)}
 =O\!\left(\frac{2^k}{k}\right).
\]
Here $(n)_k=n(n-1)\cdots(n-k+1)$, and the bound is uniform because $k=O(\log n)$. Thus $\mathbb E\sum_{k=3}^g kX_k=O(2^g)=O(n^{1/4})$, and Markov's inequality gives $o(n)$ short-cycle edge incidences with high probability. Conditioning on simplicity preserves this conclusion because the probability of simplicity is bounded away from zero \cite{Wormald1999}.

Delete the edges on these short cycles. The remaining graph has $3n/2-o(n)$ edges and maximum degree three. Each edge of a maximal matching accounts for at most five remaining edges, so such a matching $M$ has $\Theta(n)$ edges. Every first-edge theta for $e\in M$ contains a cycle through $e$, formed by its branch beginning with $e$ and either other branch. Its support therefore has more than $g$ vertices. In a vertex-disjoint batch there can be at most $n/(g+1)$ such supports, whereas $|M|=\Theta(n)$. At least $\Omega(g)=\Omega(\log n)$ batches are necessary.
\end{proof}

\begin{remark}
Theorem~\ref{thm:barrier} is not a lower bound on $\Rot(G_{n,3})$. It rules out constant overhead only for independent per-edge theta supports batched by vertex-disjointness. Shared global rotations, Hamilton-cycle methods, or algorithms that never compile a matching schedule layer by layer remain possible routes below $O(\log^9 n)$.
\end{remark}

\section{Vacancy-Assisted Routing on Cubic Expanders}\label{sec:vacancy}\label{app:vacancy}

With at least two empty vertices, a forbidden exchange of adjacent pebbles can be implemented using path shifts. The edge exchange operation requires holes at two local vertices. The main issue is to supply these holes to many exchanges simultaneously. This section proves a deterministic bound in terms of the number of holes, the number of edges whose neighborhoods are not tree-like, and the matching-routing number.

Write $B_R(u)$ for the set of vertices at distance at most $R$ from $u$. An edge $uv$ is \emph{$R$-good} if at least one of the subgraphs $G[B_R(u)]$ and $G[B_R(v)]$ is a tree. Let $b_R(G)$ be the number of edges that are not $R$-good. As before, $\delta(S)$ denotes the set of edges with exactly one endpoint in $S$, and $\rt(G)$ is the matching-routing number.

\begin{theorem}[Variable-hole transfer]\label{thm:variable-hole}
For every $\phi>0$ there are constants $R=R(\phi)$ and $C=C(\phi)$ with the following property. Let $G$ be a cubic graph on $n$ vertices such that
\begin{enumerate}
    \item $|\delta(S)|\ge\phi|S|$ for every nonempty $S\subseteq V(G)$ with $|S|\le n/2$;
    \item $G-\{u,v\}$ is connected for every edge $uv$.
\end{enumerate}
For $p$ labeled pebbles and $h=n-p\ge2$ holes,
\begin{equation}\label{eq:variable-hole-bound}
    \Mix_p(G)\le C\,\rt(G)
    \left[1+\frac ph+\log(\min\{p,h\}+2)+\min\{b_R(G),p\}\right].
\end{equation}
The bound is witnessed by schedules consisting only of path shifts.
\end{theorem}

Each term in the bracket describes one part of the construction. The ratio $p/h$ counts batches when the number of simultaneous exchanges is limited by the available holes. The logarithm accounts for repeatedly completing a constant fraction of the exchanges in a final batch. The term involving $b_R(G)$ counts sequentially processing the rare edges whose neighborhoods contain short cycles. The constant term includes the single round that handles all pebble--hole edges of a matching. We prove the local edge exchange operation first, then the parallel delivery lemma, and finally give the bound on one round of the matching-routing schedule.

\subsection{Local path-shift operations}

A \emph{path shift} acts on a path whose last vertex is empty and whose other vertices are occupied. Write the path as $(v_0,v_1,\ldots,v_\ell)$. Each pebble moves from $v_i$ to $v_{i+1}$ in one synchronous round. Its reverse is the path shift on $(v_\ell,\ldots,v_0)$ in the resulting configuration. The length of the path does not affect the number of rounds: each pebble traverses only one edge.

The two-vacancy exchange at a branching vertex and the strategy of clearing its neighbors, exchanging pebbles, and reversing the path shifts appear in Push and Swap~\cite[Section~3.1 and Figure~1]{LunaBekris2011PushSwap}. We record the exchange in our synchronous model, where it takes three path shifts. The following delivery lemma supplies these vacancies to many exchanges simultaneously.

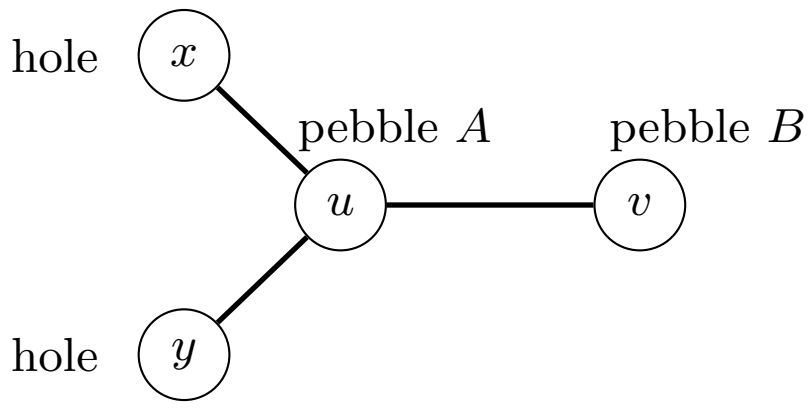
\begin{figure}[t]
\centering
\begin{tikzpicture}[scale=0.9, every node/.style={circle,draw,minimum size=6mm,inner sep=1pt,font=\small}, >=Latex]
  \node (u) at (0,0) {$u$};
  \node (v) at (2.2,0) {$v$};
  \node (x) at (-1.15,1.1) {$x$};
  \node (y) at (-1.15,-1.1) {$y$};
  \draw[line width=1pt] (u)--(v);
  \draw[line width=1pt] (u)--(x);
  \draw[line width=1pt] (u)--(y);
  \node[draw=none,rectangle] at (2.70,0.55) {pebble $B$};
  \node[draw=none,rectangle] at (0.40,0.55) {pebble $A$};
  \node[draw=none,rectangle] at (-2.1,1.1) {hole};
  \node[draw=none,rectangle] at (-2.1,-1.1) {hole};
\end{tikzpicture}
\caption{A T-junction at the cubic vertex $u$. The vertices $x,y$ are the two neighbors of $u$ other than $v$. Two holes at $x,y$ allow a three-round exchange across $uv$. The delivery lemma supplies such pairs of holes in parallel while keeping the occupied edge endpoints fixed.}
\label{fig:t-junction}
\end{figure}

\begin{lemma}[T-junction word]\label{lem:t-junction}
Let $uv$ be an occupied edge in a cubic graph, and let $x,y$ be the other two neighbors of $u$ (Figure~\ref{fig:t-junction}). If $x,y$ are holes, the pebbles at $u,v$ can be exchanged in three rounds of path shifts. The holes are returned to $x,y$, and every other pebble is fixed.
\end{lemma}

\begin{proof}
Execute the path shifts
\[
    v\to u\to x,
    \qquad x\to u\to y,
    \qquad y\to u\to v.
\]
Writing the occupancies in the order $(u,v,x,y)$ gives
\begin{align*}
 (A,B,\varnothing,\varnothing)
 &\longrightarrow(B,\varnothing,A,\varnothing)\\
 &\longrightarrow(A,\varnothing,\varnothing,B)
 \longrightarrow(B,A,\varnothing,\varnothing).
\end{align*}
Every arrow is a path shift ending at a hole, so every round is legal.
\end{proof}

\begin{lemma}[Two-hole placement]\label{lem:two-hole}
Let $D$ be a connected graph containing at least two holes. For any distinct $x,y\in V(D)$, at most three path shifts make both $x,y$ holes. The sequence is reversible.
\end{lemma}

\begin{proof}
If $x$ is occupied, shift along a shortest path from $x$ to a hole. By minimality, every internal vertex is occupied, so this delivers a vacancy to $x$ in one round. If $y$ is already empty there is nothing more to do. Otherwise let $C$ be the component of $D-x$ containing $y$. If $C$ contains a hole, shifting along the shortest path inside $C$ from $y$ to that hole delivers a vacancy to $y$ without touching $x$.

If $C$ contains no hole, shift from $y$ to the hole at $x$ along a shortest path, emptying $y$ and filling $x$. A second hole lies outside $C$. Every component of $D-x$ is adjacent to $x$, so $D-C$ is connected and contains both $x$ and that hole. A shortest path inside $D-C$ therefore delivers a vacancy to $x$ without touching $y$. At most three shifts were used in total. Reversing their order and directions reverses the construction.
\end{proof}

\begin{corollary}[Sequential fallback]\label{cor:fallback}
Let $G$ be cubic. Suppose $uv$ is occupied, $G-\{u,v\}$ is connected, and at least two holes are present. The pebbles at $u,v$ can be exchanged in at most nine path-only rounds while restoring the hole set and every other pebble.
\end{corollary}

\begin{proof}
Apply Lemma~\ref{lem:two-hole} inside $D=G-\{u,v\}$ to deliver holes to the two neighbors of $u$ other than $v$ in at most three rounds. Apply Lemma~\ref{lem:t-junction}, then reverse the hole delivery. The exchange does not change any pebble or hole inside $D$ after the placement, so the reverse shifts are legal and restore it exactly.
\end{proof}

\begin{proposition}[Bridgeless does not imply 3-connected]\label{prop:not3conn}
There are connected bridgeless cubic graphs with an edge $uv$ such that $G-\{u,v\}$ is disconnected.
\end{proposition}

\begin{proof}
Take two copies $H_1,H_2$ of $K_4$, deleting one edge $a_ib_i$ in each. Add vertices $u,v$ and the edges $uv$, $ua_1$, $ua_2$, $vb_1$, $vb_2$. The resulting simple graph is cubic. Every edge inside an $H_i$ lies on a cycle, and a path from $a_i$ to $b_i$ inside $H_i$, together with $ua_i$, $b_iv$, and $vu$, places each new edge on a cycle. Since all edges lie on cycles, the graph is bridgeless. Deleting the vertices $u,v$ separates $H_1$ from $H_2$.
\end{proof}

Thus bridgelessness alone is too weak for the sequential fallback. Theorem~\ref{thm:variable-hole} explicitly assumes that deleting the endpoints of any edge leaves a connected graph. In particular, 3-vertex-connectivity suffices.

\subsection{Good edges, ports, and private branches}

For each $R$-good edge $e=uv$, choose an endpoint $u$ whose radius-$R$ ball induces a tree. We regard $u$ as the \emph{root} of the edge. The two neighbors of $u$ other than $v$ are its \emph{ports}. The outward subtree rooted at either port, restricted to distance at most $R$ from $u$, is its \emph{private branch}. Because the graph is cubic and the induced ball is a tree, each branch has
\[
    L_R=1+2+\cdots+2^{R-1}=2^R-1
\]
vertices. The two private branches are disjoint and avoid $u,v$.

Build an auxiliary conflict graph whose vertices are the $R$-good edges. Join two conflict graph vertices when their corresponding radius-$R$ balls intersect. The two roots have distance at most $2R$. A cubic graph has only $O(2^{2R})$ vertices within that distance of any root, and a vertex is the root of at most three edges. Thus the conflict graph has maximum degree $O_R(1)$. Greedy coloring gives $Q_R=O_R(1)$ \emph{neighborhood colors}. Within one color, the balls, their ports, and their private branches are pairwise disjoint.

\subsection{Delivering holes to a constant fraction of a batch}

\begin{lemma}[Vacancy-delivery phase]\label{lem:delivery}
Fix $\phi>0$. There exist constants $R=R(\phi)$ and $c=c(\phi)>0$ such that the following holds. Suppose $G$ is cubic and $|\delta(S)|\ge\phi|S|$ whenever $0<|S|\le n/2$. Let $F$ be a matching of $m$ occupied, $R$-good edges of one neighborhood color, and let $h$ be the number of holes. If $m\le ch$, then at most five path-only rounds exchange the pebbles on at least $3m/4$ edges of $F$, restore the entire hole set, and restore every pebble not on a completed edge.
\end{lemma}

\begin{proof}
The edge-expansion hypothesis implies
\[
    |N(S)\setminus S|\ge\alpha|S|,
    \qquad \alpha=\phi/3,
    \qquad 0<|S|\le n/2,
\]
since an outside vertex is incident to at most three cut edges. Choose $R$ with $K=\alpha(2^R-1)\ge15$, and set $c=\min\{1/2,\phi/24\}$. The assertion is immediate for $m=0$, so assume $m>0$.

Let $Z$ be the $2m$ occupied edge endpoints and $T$ the $2m$ ports. These sets are disjoint because their radius-$R$ balls are disjoint. Write $H$ for the hole set, $Q=T\cap H$ for the already empty ports, and $A=T\setminus Q$ for the occupied ports. Put $q=|Q|$ and $a=|A|=2m-q$, and let $B=H\setminus Q$. Since $m\le h/2$,
\[
    |B|=h-q\ge2m-q=a.
\]
If $a=0$, all port pairs are already empty and the three-round local operations suffice. Otherwise introduce $G_0=G-(Z\cup Q)$. Deleting $Z$ protects the pebbles to be exchanged, and deleting $Q$ preserves the ports already empty.

Let $k$ be the maximum number of pairwise vertex-disjoint $A$--$B$ paths in $G_0$, including disjoint endpoints, and put $d=a-k$. Thus $d$ counts the occupied ports left unmatched by a maximum path family. By the vertex form of Menger's theorem there is a set $S_0\subseteq V(G_0)$ of size $k$ separating $A\setminus S_0$ from $B\setminus S_0$. The separator is allowed to contain vertices belonging to $A$ or $B$. Let $C$ be the union of the components of $G_0-S_0$ meeting $A\setminus S_0$.

The $a$ occupied ports have pairwise disjoint private branches. These branches avoid $Z\cup Q$: each lies in its own edge's ball, avoids that edge's endpoints, and is disjoint from its other port's branch. At most $k$ branches meet $S_0$. Every branch avoiding $S_0$ lies wholly in a component counted in $C$, giving
\begin{equation}\label{eq:C-lower}
    |C|\ge L_R d.
\end{equation}
Moreover, $C$ has no holes. All holes belong to $Q\cup B$; $Q$ was deleted, $B\setminus S_0$ is separated from $C$, and $C$ avoids $S_0$.

Every outside neighbor of $C$ belongs to $Z\cup Q\cup S_0$, whose size is
\[
    2m+q+k=4m-d\le4m.
\]
We first show that $|C|\le n/2$. Otherwise $U=V(G)\setminus C$ has size below $n/2$ and contains all $h$ holes. Cubicity and the bound above give
\begin{align*}
    \phi h&\le\phi|U|\le|\delta(U)|=|\delta(C)|\\
    &\le3(4m-d)\le12m\le12ch\le\phi h/2,
\end{align*}
a contradiction. Vertex expansion now applies to $C$ and, with \eqref{eq:C-lower}, yields $Kd\le4m-d$. This inequality also holds if $C$ is empty, since then $d=0$. Consequently,
\begin{equation}\label{eq:d-upper}
    d=a-k\le\frac{4m}{K+1}\le\frac m4.
\end{equation}

Choose $k$ disjoint $A$--$B$ paths. On each, retain the segment from its last vertex in $A$ to its first subsequent vertex in $B$. These segments remain disjoint. Their interiors avoid $A\cup B$, and all their vertices avoid $Z\cup Q$. Every interior vertex is therefore occupied, and each terminal vertex is a hole. Shift all these paths simultaneously in one round. This creates holes at $k$ additional ports while preserving the $q$ ports already empty.

Exactly $d$ ports remain occupied, so at most $d$ edges fail to have both ports empty. By \eqref{eq:d-upper}, at least $m-d\ge3m/4$ T-junctions are ready. They have disjoint vertex sets and execute simultaneously in three rounds. Every used port is again empty afterward. Moreover, the local operations fix every other vertex of every delivery path: its interior avoids $T\cup Z$, and its terminal vertex lies in $B$, disjoint from $T\cup Z$. Thus the entire configuration on each delivery path agrees with its configuration immediately after delivery. Reversing all delivery paths in one final round restores the original holes and all pebbles except those on the completed edges. The total is $1+3+1=5$ rounds.
\end{proof}

\subsection{Compiling a matching layer}

For a matching $M$ and a pebble configuration $\sigma$, define $\tau_M(\sigma)$ as the resulting configuration by acting independently on its edges. Exchange the labels on a pebble--pebble edge, slide the pebble across a pebble--hole edge, and do nothing on a hole--hole edge. Equivalently, for a hole--hole edge, assign distinct dummy labels to the holes, perform one matching-routing round, and forget the dummy identities.

\begin{lemma}[One matching layer]\label{lem:layer}
Under the hypotheses of Theorem~\ref{thm:variable-hole}, every matching action $\tau_M$ can be implemented in
\[
    O_\phi\!\left(1+\frac ph+\log(\min\{p,h\}+2)
       +\min\{b_R(G),p\}\right)
\]
path-only rounds.
\end{lemma}

\begin{proof}
First process the pebble--pebble edges, leaving the other matching edges to the end. Process the good edges by neighborhood color. For one color, let $m$ be its initial number of active edges. Let $s=\lfloor ch\rfloor$ be the largest batch accepted by Lemma~\ref{lem:delivery}. Suppose first that $h$ exceeds a constant depending only on $c$, so $s>0$. While more than $s$ edges remain, apply Lemma~\ref{lem:delivery} to any $s$ of them. Each phase completes at least $3s/4=\Theta_\phi(h)$ exchanges, so the number of these phases is $O_\phi(m/h)$.

When at most $s$ edges remain, apply the lemma to all remaining edges. Each phase leaves at most one quarter of them unprocessed. The number of phases in this final stage is
\[
 O(\log(\min\{m,h\}+2)).
\]
Each phase fixes every pebble outside its completed edges and restores the hole set. A later phase may temporarily move pebbles on earlier completed edges, but restores them before it ends. If $h$ is below the constant threshold instead, use Corollary~\ref{cor:fallback} sequentially, at a cost of $O(m)=O_\phi(m/h)$ rounds because $2\le h=O_\phi(1)$.

There are $Q_R=O_\phi(1)$ colors and at most $p/2$ pebble--pebble edges in total. Thus all good edges cost
\[
    O_\phi\!\left(1+\frac ph+\log(\min\{p,h\}+2)\right)
\]
rounds. At most $\min\{b_R(G),p/2\}$ active edges are bad, and Corollary~\ref{cor:fallback} handles these sequentially. Every operation so far restores the hole set and fixes all labels outside the exchanged edges. Therefore the pebble--hole edges still have their original occupancies and slide simultaneously in one final round. The hole--hole edges require no action.
\end{proof}

\begin{proof}[Proof of Theorem~\ref{thm:variable-hole}]
Give the $h$ holes distinct dummy labels and assign them bijectively to the target holes. The real and dummy labels define a permutation on $V(G)$. By the definition of $\rt(G)$, this permutation has a matching-routing schedule of at most $\rt(G)$ layers. Lemma~\ref{lem:layer} implements the projected action of each layer after dummy identities are forgotten. This projection depends only on the real labels and the hole set, so the physical schedule need not preserve the identities assigned to the holes. Induction over the layers yields the target configuration and the bound \eqref{eq:variable-hole-bound}.
\end{proof}

\paragraph{Constructing the schedule}
For fixed $\phi$, the radius $R$ is constant, so good edges and neighborhood colors can be found in $O_\phi(n)$ time. Each delivery phase uses a flow network with $O(n)$ arcs, obtained by splitting every vertex, including path endpoints, into an incoming and an outgoing copy joined by a capacity-one arc. A batch of $b$ edges needs at most $2b$ augmentations, each found in $O(n)$ time, and at most one final unsuccessful search. Across all phases for a color containing $m$ active edges, the sum of the batch sizes is at most $4m/3$: each phase completes at least three quarters of its batch, and each edge is completed once. Thus all flow computations for that color take $O(mn)$ time. Breadth-first searches implement each sequential fallback in $O(n)$ time. Recording the paths and updating the configuration also take $O(n)$ time per phase. Consequently, given a virtual schedule with $t$ layers, a direct implementation constructs its path-shift compilation in $O_\phi(tn^2+n)$ time. This bound includes writing the path lists. Obtaining the virtual matching schedule is a separate task.

\subsection{Random cubic graphs and the number of holes}

We now expose the random-graph estimates needed in \eqref{eq:variable-hole-bound}. Wormald's connectivity theorem gives 3-vertex-connectivity of $G_{n,3}$ with high probability \cite{Wormald1981}. Friedman's theorem gives, with high probability, a fixed bound $\lambda_*<3$ on the absolute values of its nontrivial adjacency eigenvalues \cite{Friedman2008}. The spectral cut inequality then gives
\[
    |\delta(S)|\ge(3-\lambda_*)\frac{|S|(n-|S|)}n
    \ge\frac{3-\lambda_*}{2}|S|\qquad (|S|\le n/2).
\]
Thus one can fix $\phi>0$, and then $R=R(\phi)$, independently of $n$. On the same event, the expander routing theorem of Alon et~al.~\cite{AlonChungGraham1994} gives $\rt(G_{n,3})=O(\log^2 n)$.

\begin{lemma}[Fixed-radius defects]\label{lem:local-defects}
For every fixed $R$, $\mathbb E b_R(G_{n,3})=O_R(1)$. In particular, $b_R(G_{n,3})=O_{\mathbb P}(1)$, and $b_R(G_{n,3})\le\log\log n$ with high probability.
\end{lemma}

\begin{proof}
Let $X_j(G)$ count the simple cycles of length $j$. If $G[B_R(u)]$ is not a tree, a non-tree edge in a breadth-first search tree yields a simple cycle of length at most $2R+1$. The vertex $u$ lies within distance $R$ of this cycle. In a cubic graph every radius-$R$ ball has at most $3\cdot2^R-2$ vertices. Therefore, writing $W_R$ for the vertices whose radius-$R$ balls are not trees,
\[
    |W_R|\le(3\cdot2^R-2)\sum_{j=3}^{2R+1}jX_j(G).
\]
Every bad edge is incident with a vertex of $W_R$, so
\begin{equation}\label{eq:bad-edge-cycle-bound}
    b_R(G)\le3(3\cdot2^R-2)\sum_{j=3}^{2R+1}jX_j(G).
\end{equation}
For each fixed $j$, the short-cycle estimate of McKay et~al.~\cite[eq.~(2.2)]{MWW2004} specializes to
\[
    \mathbb E X_j(G_{n,3})=\frac{2^j}{2j}\bigl(1+O_j(n^{-1})\bigr).
\]
Taking expectations in \eqref{eq:bad-edge-cycle-bound} proves the first assertion. Markov's inequality gives tightness and, explicitly,
\[
    \Pr\bigl(b_R(G_{n,3})>\log\log n\bigr)
       \le\frac{O_R(1)}{\log\log n}=o(1).
\]
\end{proof}

\begin{corollary}[Arbitrary hole count]\label{cor:rc-variable}
For even $n$, with high probability, simultaneously for every $p$ with $h=n-p\ge2$,
\[
    \Mix_p(G_{n,3})=
    O\!\left(\left[1+\frac ph+\log(\min\{p,h\}+2)
                    +\log\log n\right]\log^2 n\right).
\]
The schedules witnessing this bound consist only of path shifts.
\end{corollary}

\begin{proof}
With high probability, G is 3-connected, has a fixed spectral gap, and satisfies $b_R(G_{n,3})\le\log\log n$. The first two give the hypotheses of Theorem~\ref{thm:variable-hole} and the bound $\rt(G_{n,3})=O(\log^2 n)$. The third controls its exceptional-edge term. All these are properties of the graph alone, so the deterministic theorem applies on this event simultaneously for all hole counts.
\end{proof}

\begin{corollary}[Fixed density on random cubic graphs]\label{cor:fixed-density}
For every fixed $0<\rho<1$, put $p=\lfloor\rho n\rfloor$. For even $n$, with high probability,
\[
    \Mix_{\lfloor\rho n\rfloor}(G_{n,3})=O_\rho(\log^3 n).
\]
\end{corollary}

\begin{proof}
Here $h=n-\lfloor\rho n\rfloor\ge(1-\rho)n$, so $p/h\le\rho/(1-\rho)$, while $\log(\min\{p,h\}+2)=O(\log n)$. The bracket in Corollary~\ref{cor:rc-variable} is therefore $O_\rho(\log n)$. Multiplying by the $O(\log^2 n)$ matching-routing bound gives the claimed exponent.
\end{proof}

The different upper bounds at full and fixed density do not establish a separation between the optimal makespans. They show that holes permit a more efficient implementation of the matching-routing schedule used here. The logarithmic counting lower bound does not settle optimal makespan for random cubic graphs at either density.

\begin{corollary}[Every density]\label{cor:all-density}
For even $n$, with high probability, simultaneously for every $0\le p\le n$,
\[
    \Mix_p(G_{n,3})=O(\log^9 n).
\]
For $h=n-p\ge2$, the upper bound of Corollary~\ref{cor:rc-variable} also applies. On the same event, uniformly for $1\le p\le n$,
\[
    \Mix_p(G_{n,3})=\Omega(\log n).
\]
\end{corollary}

\begin{proof}
The first bound follows from density monotonicity (Proposition~\ref{prop:monotonicity}) and Corollary~\ref{cor:random-cubic}. Intersect that event with the one in Corollary~\ref{cor:rc-variable} for the second bound. Because $\Mix_p$ is finite, every configuration is reachable from every other, so Proposition~\ref{prop:counting} applies with maximum degree three and gives the uniform lower bound for every $p\ge1$.
\end{proof}
\section{Planar Makespan Bounds}\label{sec:planar}

Planarity interacts with the cycle structure used by a routing schedule. The quadratic one-chord witness of Theorem~\ref{thm:chorded-lower} is planar, whereas simplicity, cubicity, and 3-connectivity together allow each matching layer to be simulated at a constant cost. This gives a tight linear class maximum at every positive pebble count.

\begin{theorem}[Planar cubic class maximum]\label{thm:planar-cubic}
For every even $n\ge4$ and every $1\le p\le n$,
\[
    \max_{\substack{|V(G)|=n\\G\text{ simple, 3-connected,}\\
    \text{planar, and cubic}}}
    \Mix_p(G)=\Theta(n).
\]
In particular, the same class maximum holds for $\Rot(G)=\Mix_n(G)$.
\end{theorem}

\subsection{The facial-star compiler}

\begin{figure}[t]
\centering
\begin{tikzpicture}[scale=0.68, rotate=90, every node/.style={circle,fill=black,inner sep=1.5pt}]
  \coordinate (a) at (90:1.55);
  \coordinate (b) at (18:1.55);
  \coordinate (c) at (-54:1.55);
  \coordinate (d) at (-126:1.55);
  \coordinate (e) at (162:1.55);
  \draw[line width=1pt] (a)--(b)--(c)--(d)--(e)--cycle;
  \coordinate (ab1) at ($(a)!0.5!(b)+(0,1.05)$);
  \coordinate (ab2) at ($(a)!0.5!(b)+(0.9,0.55)$);
  \draw[line width=1pt] (a)--(ab1)--(ab2)--(b);
  \coordinate (cd1) at ($(c)!0.5!(d)+(0,-1.05)$);
  \coordinate (cd2) at ($(c)!0.5!(d)+(-0.9,-0.55)$);
  \draw[line width=1pt] (c)--(cd1)--(cd2)--(d);
  \node[draw=none,fill=none] at (0,0) {$F$};
  \node[draw=none,fill=none] at (1.0,1.4) {$F_1$};
  \node[draw=none,fill=none] at (-0.2,-1.6) {$F_2$};
\end{tikzpicture}
\caption{A facial star. All matched central-face edges are transposed by one simultaneous round on the leaf faces, one on the modified boundary $C'$, and one on the central face $F$.}
\label{fig:facial-star}
\end{figure}
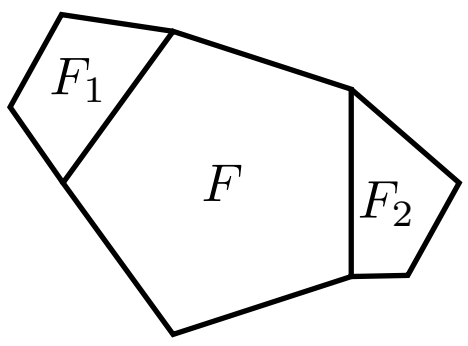

In a simple 3-connected plane graph, face boundaries are simple cycles and two distinct face closures meet in at most one vertex or one edge. In the cubic case, any two faces meeting at a vertex share an edge through that vertex: the three sectors at the vertex are pairwise consecutive. Thus face boundaries are vertex-disjoint exactly when their dual vertices are nonadjacent. These facts supply both the local rotations and the disjointness test for the compiler (Figure~\ref{fig:facial-star}).

\begin{lemma}[Facial-star identity]\label{lem:facial-star}
Let $G$ be a simple 3-connected planar cubic graph, let $F$ be a face, and let $e_i=u_iv_i\subseteq\partial F$ be matching edges. Let $F_i$ be the other face incident to $e_i$, and assume the faces $F_i$ are pairwise vertex-disjoint. Then all transpositions $(u_i\ v_i)$ can be implemented simultaneously in three rotation rounds supported on $F\cup\bigcup_iF_i$: one round rotating all leaf faces $\partial F_i$, one rotating the modified boundary $C'$ obtained from $\partial F$ by replacing each $e_i$ with the complementary path of $\partial F_i$, and one rotating $\partial F$.
\end{lemma}

\begin{proof}
Write $C=\partial F$ and $\partial F_i=e_i\cup P_i$, where $P_i$ is the complementary $u_i$--$v_i$ path along $\partial F_i$. Since two faces share at most one edge and, in a cubic plane graph, faces meeting at a vertex share an edge through it, $\partial F_i\cap C=e_i$, and $P_i$ has at least one internal vertex. Pairwise vertex-disjointness of the $F_i$ makes the paths $P_i$ mutually disjoint. Replacing each $e_i$ in $C$ by $P_i$ therefore produces a simple cycle $C'$.

Choose an orientation of $C$ and name the endpoints of each $e_i$ so that $C$ traverses it as $u_i\to v_i$. Orient $C'$ compatibly, traversing each $P_i$ from $u_i$ to $v_i$, and orient $\partial F_i$ along $P_i$ from $u_i$ to $v_i$ and back across $e_i$. Applying the rightmost operation first, the identity is
\[
    R_C\,R_{C'}^{-1}\left(\prod_i R_{\partial F_i}\right)
    =\prod_i(u_i\ v_i).
\]
The leaf rotations run simultaneously because their boundaries are vertex-disjoint. For $P_i=(u_i,w_1,\ldots,w_\ell,v_i)$, the first two rounds restore every interior label of $P_i$. They move the label of $u_i$ to $w_1$ and back to $u_i$, and move the label of $v_i$ first to $u_i$ and then to the predecessor of $u_i$ on $C$. The last assertion uses that the edges $e_i$ form a matching, so the edge preceding $u_i$ is not replaced. Every other label on $C$ moves to its predecessor in the first two rounds. The final rotation of $C$ therefore exchanges each pair $u_i,v_i$ and restores every other label.
\end{proof}

We use an elementary decomposition in place of an optimized star-arboricity bound. A \emph{star forest} is a forest whose nontrivial components are stars. In particular, its different stars have disjoint vertex sets.

\begin{lemma}[Elementary planar batching]\label{lem:planar-star-decomposition}
The edges of a simple planar graph can be partitioned into ten star forests. Such a partition can be found in linear time.
\end{lemma}

\begin{proof}
Euler's formula implies that every nonempty subgraph has a vertex of degree at most five. Successively delete such a vertex and orient its remaining incident edges toward the undeleted vertices. The resulting orientation is acyclic and has outdegree at most five. At each vertex, give its outgoing edges distinct indices in $\{1,\ldots,5\}$. The edges of any one index form a forest: an undirected cycle would either have a vertex with two outgoing edges or be a directed cycle, both impossible. Root each tree of each forest and split its edges according to the parity of the depth of their parent endpoint. Each part is a star forest, giving ten in total. Degree buckets, tree searches, and parity assignments implement these steps in linear time.
\end{proof}

\begin{proof}[Proof of Theorem~\ref{thm:planar-cubic}, upper bound]
Fix a matching $M$ of a matching-routing schedule and pass to the planar dual $G^*$, where each edge of $M$ joins its two incident faces. The dual is simple and planar. By Lemma~\ref{lem:planar-star-decomposition}, the dual edges corresponding to $M$ partition into ten star forests. Fix one forest. Each star has a center face $F$ and leaf faces $F_i$, and its star edges are matched edges on $\partial F$.

Properly color $G^*$ with six colors by processing a degree-at-most-five deletion order backward. In each star, the leaves avoid the center's color and fall into at most five color classes. Leaves of one color are nonadjacent in $G^*$, so their face boundaries are pairwise vertex-disjoint in $G$. Order each star's nonempty leaf classes arbitrarily and process its $j$th class in stage $j$, for at most five stages.

Different stars of one stage may still touch in the primal graph. In the dual, however, each center together with its current leaves is a connected vertex set, and these sets are disjoint because they belong to different components of a star forest. Contract all these sets and delete the unused dual vertices. If two corresponding primal supports share a vertex, some face in one support and some face in the other meet there. Cubicity makes those faces adjacent in the dual, so the two contracted vertices are adjacent. The support-conflict graph is therefore a subgraph of this planar minor after loops and parallel edges are removed. A further six-coloring makes the primal supports within each batch vertex-disjoint.

Altogether one matching layer splits into at most $10\cdot5\cdot6=300$ batches of vertex-disjoint facial stars. Lemma~\ref{lem:facial-star} executes each batch in three rounds, so
\[
    \Rot(G)\le900\,\rt(G).
\]
Since every connected $n$-vertex graph has $\rt(G)<3n$ \cite[Theorem~1]{AlonChungGraham1994},
\[
    \Rot(G)=O(n)
\]
for every simple 3-connected planar cubic graph, and Proposition~\ref{prop:monotonicity} extends the upper bound to $\Mix_p(G)$ for every $p$.
\end{proof}

The constant is deliberately unoptimized. The five-star-forest theorem of Hakimi, Mitchem, and Schmeichel~\cite{HakimiMitchemSchmeichel1996} would halve it. The elementary proof above suffices for the linear bound. Given a plane embedding and a $t$-layer virtual schedule, all facial boundaries, dual colorings, and batch assignments can be constructed in $O(n)$ preprocessing and $O(n)$ time per layer. Indeed, the total length of all face boundaries is $2|E(G)|=3n$, and every stage scans only a constant number of such boundary lists. Thus this planar conversion takes $O(n(t+1))$ time, including its explicit move lists.

\begin{proof}[Proof of Theorem~\ref{thm:planar-cubic}, lower bound]
For $m\ge3$, the prism $C_m\square K_2$ is simple, planar, cubic, and 3-connected, has $n=2m$ vertices, and has diameter $\Theta(n)$. The full-occupancy upper bound just proved, together with Proposition~\ref{prop:monotonicity}, guarantees reachability for every smaller pebble count. Proposition~\ref{prop:dist} therefore gives $\Mix_p(C_m\square K_2)=\Omega(n)$ for every $p\ge1$, already with a single pebble placed across a diametral pair. This covers every even $n\ge6$. The remaining case $n=4$ is $K_4$, where any nontrivial instance needs a round, and $\Theta(n)$ absorbs finitely many cases.
\end{proof}

The hypotheses are substantive. Cubicity alone does not imply connectivity; connected bridgeless cubic graphs are 2-vertex-connected but need not be 3-connected (Proposition~\ref{prop:not3conn} in Section~\ref{sec:vacancy}). The facial argument uses arbitrary simple-cycle rotations, since the modified boundary $C'$ need not be a face. A facial-rotations-only model is therefore not covered by this proof.

\subsection{Recovered lattice benchmarks}\label{app:lattice}

The same transfer principle recovers known bounds on square and triangular lattice patches. We include the derivation to illustrate the bounded-support case. These asymptotic orders are established benchmarks.

\begin{proposition}[Bounded local supports]\label{prop:bounded-support}
Let $H$ be a spanning subgraph of $G$. Suppose every edge $e\in E(H)$ lies in a 2-connected noncycle subgraph $K_e\subseteq G$ with at most $s$ vertices, and every vertex of $G$ lies in at most $\ell$ selected supports, where $s,\ell$ are absolute constants. Then $\Rot(G)=O(\rt(H))$.
\end{proposition}

\begin{proof}
By Yu and Rus~\cite{YuRus2015}, the endpoint transposition of $e$ can be implemented inside $K_e$, fixing every other support vertex, using a number of rotations depending only on $s$. The compiler argument of Theorem~\ref{thm:compiler} applies unchanged to the $\rt(H)$ layers of a matching-routing schedule on $H$ (supports are needed only for edges of $H$), with constant support size, word length, and load.
\end{proof}

\begin{proposition}[Recovered lattice benchmark]\label{prop:lattice}
Let $Q_{a,b}$ be the $a\times b$ square grid with $a,b\ge3$, and let $T_{a,b}$ be the triangular patch obtained by adding one consistent diagonal to every square cell. For $G\in\{Q_{a,b},T_{a,b}\}$ and every $1\le p\le ab$,
\[
    \Mix_p(G)=\Theta(a+b).
\]
\end{proposition}

\begin{proof}
Rectangular grids have $\rt(Q_{a,b})=O(a+b)$ \cite{AlonChungGraham1994,AlpertEtAl2022}, and $Q_{a,b}$ is a spanning subgraph of $T_{a,b}$, so the same virtual schedule is available in both. Every square-grid edge lies in a nearby $2\times3$ or $3\times2$ rectangular subgrid, which has at most six vertices, is 2-connected, and is not a cycle; the natural local choice has constant vertex load. Proposition~\ref{prop:bounded-support} gives $\Rot=O(a+b)$ for both families, and Proposition~\ref{prop:monotonicity} extends this to every $p$. Both families have diameter $\Theta(a+b)$, so Proposition~\ref{prop:dist} supplies the matching lower bound already with one pebble.
\end{proof}

The square-grid order in Proposition~\ref{prop:lattice} already follows from Demaine et~al.~\cite{DemaineEtAl2019}, and triangular-grid routing is likewise covered, under a stricter geometric collision model, by Chinta et~al.~\cite{ChintaHanYu2020}; the polyomino setting of Fekete et~al.~\cite{FeketeEtAl2024} gives a further close comparison at maximum density.
\section{Further Consequences of Local Transpositions}\label{sec:extensions}\label{app:dense}

The preceding sections concern sparse graphs. We now quantify how larger vertex connectivity and common neighborhoods supply smaller transposition supports. Connectivity provides many disjoint paths from which to select a support, and common neighbors provide constant-size supports directly. These deterministic results also yield consequences for random regular graphs.

\subsection{A Connectivity Compiler and Spectral Expanders}

We first bound the simulation cost in terms of vertex connectivity, then combine it with spectral matching-routing estimates.

\begin{theorem}[Connectivity compiler]\label{thm:conn-compiler}
Let $G$ have vertex connectivity $k\ge3$. Then
\[
    \Rot(G)=O\!\left[
      \frac nk\left(\frac nk+\log n\right)\rt(G)
    \right],
\]
and the same bound holds for $\Mix_p(G)$ at every $p$.
\end{theorem}

\begin{proof}
Fix one matching $M$ of a matching-routing schedule. For $e=uv\in M$, Menger's theorem gives $k$ internally vertex-disjoint $u$--$v$ paths, and at least $k-1$ of them avoid the edge $e$ itself. Their internal vertex sets are disjoint and total at most $n-2$ vertices, so fewer than $(k-1)/4$ of these paths have at most $4(n-2)/(k-1)$ internal vertices. Choose two of such short paths uniformly at random without replacement. Combining them with the single-edge branch $e$ they form a first-edge theta support $H_e$ of size $L=O(n/k)$, and Lemma~\ref{lem:theta} implements the transposition in three rounds.

Make these choices independently over $e\in M$. Fix a vertex $x$. Apart from the at most one matching edge incident to $x$, the vertex $x$ lies on at most one candidate path of any fixed $e$, because the candidate family is internally disjoint. Therefore $\Pr(x\in H_e)=O(1/k)$, and the expected number of chosen supports containing $x$ is $O(n/k)$. A Chernoff bound and a union bound over the $n$ vertices yield a simultaneous choice with maximum vertex load $\mu_H=O(n/k+\log n)$. The conflict graph then has maximum degree $O(L\mu_H)$, so one layer compiles into $O\bigl((n/k)(n/k+\log n)\bigr)$ rotation rounds. Multiply the number of rounds by $\rt(G)$. Density monotonicity (Proposition~\ref{prop:monotonicity}) extends the bound to $\Mix_p(G)$ for every $p$.
\end{proof}

An $(n,d,\lambda)$-graph is a $d$-regular graph on $n$ vertices whose adjacency eigenvalues $d=\lambda_1\ge\cdots\ge\lambda_n$ satisfy $\max_{2\le i\le n}|\lambda_i|\le\lambda$. Thus the bound controls both ends of the nontrivial spectrum and excludes a second eigenvalue equal to $d$.

\begin{lemma}[Spectral expansion forces connectivity]\label{lem:spectral-conn}
Every $(n,d,\lambda)$-graph with $\lambda\le d/2$ has vertex connectivity $\kappa(G)\ge d/8$.
\end{lemma}

\begin{proof}
The claim is immediate for a complete graph. Otherwise, let $S$ be a vertex separator and $A$ a smallest component of $G-S$, so $|A|\le n/2$ and every edge leaving $A$ ends in $S$. If $|A|\le d/2$, a vertex of $A$ has at least $d-|A|+1\ge d/2$ neighbors in $S$. If $|A|>d/2$, the spectral cut estimate gives
\[
    |\delta(A)|\ge(d-\lambda)\frac{|A|(n-|A|)}n\ge\frac{d|A|}4,
\]
while at most $d|S|$ edges end in $S$. Thus $|S|\ge|A|/4>d/8$.
\end{proof}

\begin{theorem}[Strong spectral expanders]\label{thm:spectral}
There are absolute constants $d_0,C$ such that every $(n,d,\lambda)$-graph with $d\ge d_0$ and $\lambda<d/72$ satisfies
\[
    \Rot(G)\le C\,\frac nd\left(\frac nd+\log n\right)\frac{\log n}{\log(d/\lambda)},
\]
and the same bound holds for $\Mix_p(G)$ at every density.
\end{theorem}

\begin{proof}
Lemma~\ref{lem:spectral-conn} gives $\kappa(G)=\Omega(d)$. Insert this into Theorem~\ref{thm:conn-compiler} and use Nenadov's bound $\rt(G)\le C'\log n/\log(d/\lambda)$, whose hypotheses are exactly $d\ge d_0$ and $\lambda<d/72$ \cite[Theorem~1.1]{Nenadov2023}.
\end{proof}

The hypothesis $\lambda<d/72$ does not apply to random cubic graphs, where $\lambda$ is near $2\sqrt2$. Nenadov's theorem does cover sufficiently strong expanders of fixed degree. It remains open in their work whether every general fixed-degree, fixed-expansion graph has $\rt(G)=O(\log n)$ \cite[Section~4]{Nenadov2023}. Horn and Purcilly~\cite{HornPurcilly2020} detail related bounds for dense and expanding graphs.

\begin{corollary}[Random regular graphs of growing degree]\label{cor:growing}
Let $d=d(n)\to\infty$, $d\le n/2$, and $nd$ even. Then \whp
\[
    \Rot(G_{n,d})=
    O\!\left[
      \frac nd\left(\frac nd+\log n\right)
      \frac{\log n}{\log d}
    \right],
\]
simultaneously for all $\Mix_p(G_{n,d})$.
\end{corollary}

\begin{proof}
Cook, Goldstein, and Johnson~\cite[Theorem~1.1]{CookGoldsteinJohnson2018} give $\lambda(G_{n,d})\le C\sqrt d$ \whp for $d\le n^{2/3}$. Tikhomirov and Youssef~\cite[Theorem~A]{TikhomirovYoussef2019}, with their parameter $\alpha=1/2$, give the same estimate for $n^{1/2}\le d\le n/2$. These overlapping ranges cover every growing degree in the statement, including arbitrarily slow growth. For sufficiently large $d$, $C\sqrt d<d/72$ and $\log(d/(C\sqrt d))=\Omega(\log d)$. Apply Theorem~\ref{thm:spectral}.
\end{proof}

The counting bound gives only $\Omega(\log n/\log(d+1))$ here. On even $n$, Proposition~\ref{prop:backbone} will give a stronger bound for a substantial part of this range by embedding a random cubic graph. The spectral theorem remains useful as a deterministic statement and has no parity restriction on $n$ beyond $nd$ being even.

\subsection{Dense Codegree}

For dense graphs, common neighbors allow transpositions of arbitrary vertex pairs on constant-size supports, avoiding a separate matching-routing schedule.

\begin{lemma}[Two involutions]\label{lem:involutions}
Following the construction of Alon, Chung, and Graham~\cite[Theorem~2]{AlonChungGraham1994}, every permutation is the product of two involutions, i.e., of two permutations whose nontrivial cycles are pairwise disjoint transpositions.
\end{lemma}

\begin{proof}
On one cycle $(v_0\ v_1\ \cdots\ v_{\ell-1})$, indices modulo $\ell$, the maps $A(v_i)=v_{-i}$ and $B(v_i)=v_{1-i}$ are involutions with $B\circ A(v_i)=v_{i+1}$. Take disjoint unions over the cycles.
\end{proof}

\begin{lemma}[$K_{2,3}$ transposition word]\label{lem:k23}
Suppose $u,v$ have three distinct common neighbors $a,b,c$. Define oriented 4-cycles $A=(u\,a\,v\,b)$, $B=(u\,a\,v\,c)$, $D=(u\,b\,v\,c)$. Then, applying the rightmost rotation first,
\[
    B\,D\,A^{-1}\,B\,A=(u\ v),
\]
and $a,b,c$ are restored. Hence any pair with three common neighbors can be transposed in five rotation rounds on a five-vertex support.
\end{lemma}

\begin{proof}
Track the five vertices through the five rotations. The composition sends the label of $u$ to $v$ and of $v$ to $u$, and returns the labels of $a,b,c$. No other vertex moves, because every rotation is supported on the displayed $K_{2,3}$.
\end{proof}

\begin{theorem}[Codegree theorem]\label{thm:codegree}
Let $q(G)=\min_{u\ne v}|N(u)\cap N(v)|$. If $q(G)\ge3$, then
\[
    \Rot(G)=O\!\left(1+\frac{n}{q(G)}\right),
\]
and consequently $\Mix_p(G)=O(1+n/q(G))$ for every $p\le n$.
\end{theorem}

\begin{proof}
By Lemma~\ref{lem:involutions} it suffices to process an involution with $m\le n/2$ arbitrary transposition pairs. Each pair has at least $q=q(G)$ candidate common neighbors. Out of these neighbors, assign three distinct helpers per pair greedily. Let $B_0$ be the per-vertex capacity, or the maximum number of pairs that any vertex can serve as a helper for.
\[
    B_0=\left\lceil\frac{3m}{q-2}\right\rceil+1.
\]
The greedy selection never fails. At most $3m$ helper assignments are made in total, so the number of saturated vertices is at most $3m/B_0<q-2$, leaving at least three unsaturated candidates for the current pair. A helper is distinct from the pair vertices because the graph is simple. Since $q/(q-2)\le3$, each vertex is a helper in at most $B_0=O(1+n/q)$ supports and a pair vertex in at most one. The five-vertex supports of Lemma~\ref{lem:k23} therefore have vertex load $O(1+n/q)$. The conflict-coloring argument of Theorem~\ref{thm:compiler} executes the involution in $O(1+n/q)$ rounds, since both support size and word length are constant. The second involution doubles the bound, and Proposition~\ref{prop:monotonicity} extends it to every $p\le n$.
\end{proof}

The proof is algorithmic. Given an adjacency matrix, compute $q(G)$ in $O(n^3)$ time. The two involutions, their helper assignments, and their conflict colorings then take $O(n^2)$ time, as does writing the schedule as one destination per vertex per round. Thus a direct implementation constructs the schedule in deterministic $O(n^3)$ time, including output.

\begin{corollary}[Minimum degree above one half]\label{cor:halfdense}
Define $\delta_{min}(G)$ as the minimum degree of $G$. If $\delta_{\min}(G)\ge(1/2+\epsilon)n$ for fixed $\epsilon>0$ and sufficiently large $n$, then $\Rot(G)=O_\epsilon(1)$ and $\Mix_p(G)=O_\epsilon(1)$ for every $p$.
\end{corollary}

\begin{proof}
Inclusion--exclusion gives $|N(u)\cap N(v)|\ge\deg(u)+\deg(v)-n\ge2\epsilon n$ for all pairs. Once $2\epsilon n\ge3$, apply Theorem~\ref{thm:codegree}.
\end{proof}

The complete graph permits an exact constant.

\begin{proposition}[Complete graphs]\label{prop:complete}
For every $n\ge4$, $\Rot(K_n)=2$, and $\Mix_p(K_n)\le2$ for every $p\le n$. ($K_3$ is a cycle, so $\Rot(K_3)=+\infty$.)
\end{proposition}

\begin{proof}
At maximum density a legal round is a vertex-disjoint union of cycle rotations of length at least three (Proposition~\ref{prop:round-decomposition}), and $K_n$ supports every such union. It therefore suffices to factor an arbitrary permutation into two permutations whose nontrivial cycles all have length at least three. Factor blockwise on disjoint vertex sets. Every cycle of length at least three is its own block, placed in the first factor. Group the 2-cycles into pairs, making one group a triple if the total number of 2-cycles is odd and at least three. Writing the rightmost factor first, a pair and a triple factor as
\begin{align*}
(a_1\,b_1)(a_2\,b_2)&=(a_1\,b_1\,a_2)\,(b_1\,a_2\,b_2),\\
(a_1b_1)(a_2b_2)(a_3b_3)&=(a_1a_2a_3)(b_1b_2b_3)(a_1b_3a_2b_1a_3b_2).
\end{align*}
If instead a single 2-cycle $(a\ b)$ remains, merge it with two fixed points via $(a\,b)=(b\,c\,d)\,(a\,d\,c\,b)$, or with a cycle of length $k\ge3$ via
\[
(a\,b)(c_1\cdots c_k)=(a\,c_1\,b\,c_2\,c_3\cdots c_k)\,(a\,c_1\,b\,c_k).
\]
For $n\ge4$ at least two vertices lie outside $\{a,b\}$, each fixed or on a cycle of length at least three, so one of the two merges is available. Each identity can be verified by tracking labels, and distinct blocks act on disjoint vertex sets, so both factors are vertex-disjoint unions of cycles of length at least three and execute in one round each. The value two is exact, since a single-transposition target moves only two labels while every nontrivial round moves at least three. Density monotonicity (Proposition~\ref{prop:monotonicity}) gives $\Mix_p(K_n)\le2$.
\end{proof}

\subsection{Connectivity without a Routing Theorem}\label{app:connectivity}

The compiler of Theorem~\ref{thm:conn-compiler} multiplies by $\rt(G)$. When no strong matching-routing bound is available, connectivity alone still gives polynomial bounds. Here the transposed vertices need not be adjacent, so the theta exchange takes time proportional to the support size rather than three rounds. All upper bounds below extend to every agent count by Proposition~\ref{prop:monotonicity}.

\begin{lemma}[Theta-terminal exchange, following Yu--Rus~\cite{YuRus2015}]\label{lem:theta-terminals}
If $H$ is a theta with terminals $u,v$, then the labels at $u,v$ can be transposed in at most $2|V(H)|+3$ rotation rounds, restoring every other label.
\end{lemma}

\begin{proof}
Let $P,Q,R$ be the three branches. Since the graph is simple, choose $P$ to have an internal vertex, and let $x$ be its first vertex after $u$. Rotate $P\cup Q$ once from $u$ into $P$. The label of $u$ now occupies $x$, and the label of $v$ occupies a vertex of $Q$, possibly $u$ itself. The cycle $Q\cup R$ avoids $x$, so at most $|V(H)|-1$ rotations of this cycle put the label of $v$ at $u$ without moving the label at $x$. Apply Lemma~\ref{lem:theta} to transpose $u,x$ in three rounds and undo all positioning rotations. This conjugation exchanges exactly the original terminal labels and uses at most $2|V(H)|+3$ rounds. It is the positioning--exchange--reverse construction in the expanded proof of \cite[Theorem~2, equation~(2)]{YuRus2015}, specialized to the terminals.
\end{proof}

\begin{theorem}[Sequential connectivity bound]\label{thm:seq-conn}
If $G$ is $k$-vertex-connected with $k\ge3$, then $\Rot(G)=O(n^2/k)$.
\end{theorem}

\begin{proof}
For any pair $u,v$, Menger's theorem gives $k$ internally vertex-disjoint $u$--$v$ paths. The three shortest have at most $3(n-2)/k$ internal vertices in total, so their union is a theta of size $O(n/k)$. Lemma~\ref{lem:theta-terminals} transposes $u,v$, restoring all other labels, in $O(n/k)$ rounds. Every permutation is a product of at most $n-1$ transpositions.
\end{proof}

\begin{theorem}[Parallel connectivity bound]\label{thm:par-conn}
For every $k$-vertex-connected $n$-vertex graph with $k\ge3$,
\[
\Rot(G)=
O\!\left(
\min\left\{
\frac{n^2}{k},\,
\frac{n^2}{k^2}\cdot\frac{\log n}{\log\log n}
\right\}
\right).
\]
\end{theorem}

\begin{proof}
The first term is Theorem~\ref{thm:seq-conn}. We prove the second for sufficiently large $k$, the other case being dominated. By Lemma~\ref{lem:involutions}, it suffices to process a set of disjoint terminal pairs.

\emph{Bulk phases.} Put $b=\lfloor k/100\rfloor$. Select $b$ unprocessed pairs $(u_i,v_i)$ and delete their $2b$ endpoints. The residual graph is at least $(k-2b)$-connected. Greedily choose three private neighbors outside the endpoint set for every $u_i$ and $v_i$, all $6b$ helpers distinct. This is possible because each terminal has at least $k$ neighbors while fewer than $8b<k$ vertices are forbidden at any step. Pair the helpers of $u_i$ with those of $v_i$. A graph is $s$-linked if any prescribed pairing of $2s$ distinct vertices can be joined by mutually vertex-disjoint paths. The residual graph is at least $98b$-connected. The theorem of Bollob\'as and Thomason~\cite{BollobasThomason1996} that $22s$-connectivity implies $s$-linkedness, with $s=3b$, therefore supplies all $3b$ helper paths. Adding the terminal--helper edges gives $b$ vertex-disjoint thetas. Their total size is at most $n$, so at least $b/2$ have size at most $2n/b=O(n/k)$. Their terminal exchanges from Lemma~\ref{lem:theta-terminals} run in parallel in $O(n/k)$ rounds. A phase thus processes $\Theta(k)$ pairs. Repeating until fewer than $b$ pairs remain costs $O(n/k)$ phases and $O(n^2/k^2)$ rounds.

\emph{Tail.} Now $m<b=O(k)$ pairs remain. For a fixed pair $(u_i,v_i)$, delete the endpoints of all other pairs. After deletion, the graph remains $\Theta(k)$-connected and contains $\Theta(k)$ internally disjoint $u_i$--$v_i$ paths avoiding all other terminals, of which a constant fraction have $O(n/k)$ vertices. Keep $t=\Theta(k)$ short paths and choose three uniformly at random for each pair independently, forming a theta $H_i$ of size $L=O(n/k)$. For a nonterminal vertex $x$, within one pair's candidate family $x$ lies on at most one path, so $\Pr(x\in H_i)=O(1/k)$; independence over the $m=O(k)$ pairs makes the expected load of $x$ constant. For a sum $X$ of independent Bernoulli variables with bounded mean, $\Pr(X\ge \tau)\le(e\,\mathbb EX/\tau)^\tau$; taking $\tau=\Theta(\log n/\log\log n)$ and union-bounding over vertices yields a choice with maximum load $\mu_H=O(\log n/\log\log n)$, while other terminals have load one by construction. Conflict coloring with $O(L\mu_H)$ colors and $O(L)$ rounds per color class costs
\[
    O(L^2\mu_H)=O\!\left(\frac{n^2}{k^2}\cdot\frac{\log n}{\log\log n}\right).
\]
The two involutions change only the constant factor.
\end{proof}

\begin{proposition}[Connectivity bottleneck]\label{prop:conn-lower}
For every $3\le k\le n/2$ there are $k$-vertex-connected $n$-vertex graphs with $\Rot(G)=\Omega(n/k)$.
\end{proposition}

\begin{proof}
Let $a=\lfloor(n-k)/2\rfloor$, $b=\lceil(n-k)/2\rceil$, and $G=K_k\vee(K_a\cup K_b)$, where $\vee$ is the graph join. Removing the central $K_k$ disconnects the side cliques, while removing fewer than $k$ vertices leaves a central vertex adjacent to everything, so $\kappa(G)=k$. Pair the $a$ vertices of the smaller side with $a$ vertices of the larger side and transpose each pair. All $2a=\Theta(n)$ crossing pebbles must occupy a central vertex at the end of some round, since the side cliques have no edges between them. A schedule of $T$ rounds provides at most $kT$ such vertex--time positions, so $T\ge2a/k=\Omega(n/k)$.
\end{proof}

The gap between $\Omega(n/k)$ and the parallel upper bound is roughly a factor $n/k$. Determining the correct dependence is open (Section~\ref{sec:discussion}).

\subsection{Random Regular Graphs via Codegrees and Subgraphs}\label{app:conditional}

Two additional random-graph inputs complement the spectral bound. For large degrees, a published codegree estimate applies directly to Theorem~\ref{thm:codegree}. For every growing degree on even orders, a monotone coupling transfers both random-cubic bounds.

\begin{theorem}[Dense random regular graphs]\label{thm:kimvu}
Let $d=d(n)$ be a positive integer with $nd$ even, $d\le n-1$, and $d^2/n=\omega(\log n)$. Then \whp
\[
    \Rot(G_{n,d})=O\!\left(1+\frac{n^2}{d^2}\right),
\]
and the same bound holds for every $\Mix_p(G_{n,d})$.
\end{theorem}

\begin{proof}
Suppose first that $d\le3n/4$. Krivelevich et~al.~\cite[Theorem~2.1(i)]{KrivelevichSudakovVuWormald2001} imply, under the stated growth condition, that \whp
\[
    \max_{u\ne v}\left||N(u)\cap N(v)|-\frac{d^2}{n}\right|
    \le C\frac{d^3}{n^2}+6d\sqrt{\frac{\log n}{n}},
\]
where the first term can be omitted when $d\ge n/\log n$. Below this threshold, its ratio to $d^2/n$ is $O(1/\log n)$; the ratio of the second term is $6\sqrt{n\log n}/d=o(1)$ throughout. Hence $q(G_{n,d})=(1+o(1))d^2/n$ \whp. Apply Theorem~\ref{thm:codegree}. For $d>3n/4$, Corollary~\ref{cor:halfdense} gives the asserted constant bound deterministically. Density monotonicity gives the same conclusion for every $p$ on the same graph event.
\end{proof}

\begin{corollary}[Linear degree]\label{cor:linear-degree}
If $d/n\to c$ for a fixed $c\in(0,1]$, along orders for which $G_{n,d}$ exists, then $\Rot(G_{n,d})=\Theta(1)$ \whp. For $c>1/2$, the same conclusion holds deterministically for every sufficiently large $d$-regular graph.
\end{corollary}

\begin{proof}
The upper bound is Theorem~\ref{thm:kimvu} or Corollary~\ref{cor:halfdense}. For the lower bound, a target consisting of a single transposition cannot be realized in one round, since every nontrivial round moves at least three labels.
\end{proof}

\begin{proposition}[Cubic-backbone transfer]\label{prop:backbone}
Let $n$ be even and $d=d(n)\to\infty$, with $d\le n-1$. Then \whp
\[
    \Rot(G_{n,d})=O(\log^9 n)
    \quad\text{and}\quad
    \Mix_p(G_{n,d})=O(\log^9 n)
\]
simultaneously for all $p$. Moreover, \whp, for every fixed $0<\rho<1$,
\[
    \Mix_{\lfloor\rho n\rfloor}(G_{n,d})=O_\rho(\log^3 n).
\]
\end{proposition}

\begin{proof}
Hollom et~al.~\cite[Theorem~1.2]{HollomEtAl2025} couple $H\sim G_{n,d_1}$ and $G\sim G_{n,d_2}$ with $H\subseteq G$ \whp whenever $n$ is even, $3\le d_1\le n^{1/7}/\log n$, $d_2\ge d_1$, and $d_2\to\infty$. Take $d_1=3$ and $d_2=d$. On the coupling event, edge monotonicity gives $\Mix_p(G)\le\Mix_p(H)$ for every $p$. Corollaries~\ref{cor:random-cubic} and~\ref{cor:fixed-density} therefore give the two bounds, with density monotonicity extending the first to all $p$.
\end{proof}

For even $n$, one may take the minimum of the applicable spectral, codegree, and cubic-subgraph bounds. The codegree construction is explicit once the graph is given; the coupling proposition asserts the existence of a suitable cubic subgraph and does not supply an efficient method for finding one in an input graph.

\section{Construction of Schedules and Open Questions}\label{sec:discussion}

The principal upper bounds combine a matching-routing schedule with a legal implementation of each layer. Cycle covers control the overlap of local supports, expansion permits simultaneous vacancy delivery, and planar faces allow several exchanges to share a rotation. The common-neighbor extension instead implements two arbitrary involutions directly. These distinctions also determine the inputs required by the proofs.

\subsection{Makespan and the cost of constructing a schedule}

The makespan counts parallel motion rounds, not the computation used to find them. The transfer proofs explicitly specify legal schedules once their routing schedules and auxiliary combinatorial objects have been supplied. Their output can be represented by a list of pebbles' moves in each round, or more compactly by the paths and cycles rotated in that round.

For the local-support transfer, suppose a virtual matching layer has $q$ edges and the chosen rotation words have supports of size at most $L$. Form a conflict graph as follows. For each vertex, list the supports incident to it, and add an edge between two vertices when the intersection of their lists is nonempty. If $d_v$ is the number of chosen supports containing $v$, constructing the conflict graph takes
\[
 O\!\left(n+qL+\sum_{v\in V}d_v^2\right)
\]
elementary operations using incidence lists. Duplicate conflict edges can be discarded. The layer schedule is then produced with greedy coloring and simultaneous execution of the words in each color. The bound on the number of colors in Theorem~\ref{thm:compiler} controls makespan, while these incidence counts describe the work of the translation. Given a cycle cover, the local trees and rotation words in Theorem~\ref{thm:cycle-cover} are obtained by graph searches. Finding the input cover and finding the virtual matching-routing schedule are additional tasks, governed by the respective external results.

The vacancy transfer is also effective. Its disjoint delivery paths can be obtained by an integral maximum-flow computation. Replace each vertex with an input and an output vertex joined by a capacity-one arc. Encode the port and hole sets as a source and a sink vertex. Each phase completes a constant fraction of the designated exchanges, and the proofs bound the number of phases. Neighborhood colorings, the sequential fallback, and the reversal of delivery paths are direct finite graph operations. The construction analysis in Section~\ref{sec:vacancy} gives $O_\phi(tn^2+n)$ computational overhead for an explicitly supplied $t$-layer matching schedule, with the expansion parameter $\phi$ fixed. This statement concerns the transfer, and does not identify its running time with its much smaller number of parallel motion rounds.

For the planar transfer, the star-forest decomposition becomes part of the elementary construction. Lemma~\ref{lem:planar-star-decomposition} finds it by degree deletion and tree searches. Given a plane embedding and a $t$-layer virtual schedule, Section~\ref{sec:planar} gives $O(n(t+1))$ conversion time, including output. The common-neighbor theorem needs no virtual schedule: its proof gives a deterministic $O(n^3)$ algorithm from an adjacency matrix, again including output.

The source of an auxiliary object matters. The random-cubic argument uses the explicit routing and cycle-cover constructions cited in Section~\ref{sec:upper}. Their computation is additional to the conversion cost. In Theorem~\ref{thm:conn-compiler}, Menger paths can be found by flow, independently sampled supports have checkable loads, and rejection sampling gives an expected polynomial construction for each supplied matching layer. By contrast, the large prescribed linkage in Theorem~\ref{thm:par-conn} and the suitable cubic subgraph in Proposition~\ref{prop:backbone} are used as existence statements. Those two makespan bounds do not by themselves supply a uniform polynomial algorithm to find the required objects. This separation preserves the extremal content without making a stronger algorithmic claim than the proofs establish.

\subsection{Extremal questions}

\paragraph{Random cubic graphs and occupancy}
The first question is to close the gaps between $\Omega(\log n)$ and $O(\log^9 n)$ at full occupancy, and between $\Omega(\log n)$ and $O_\rho(\log^3 n)$ at every fixed density below one. Theorem~\ref{thm:barrier} excludes constant overhead for a compiler that assigns independent first-edge theta supports and batches only vertex-disjoint supports. It places no lower bound on the optimal makespan and does not exclude shared rotations or constructions that avoid compiling each matching layer separately. Below full occupancy, a reusable, well-distributed set of holes may reduce the repeated cost of vacancy delivery. Whether the true asymptotic worst-case makespan differs between full occupancy and any fixed smaller density remains open.

\paragraph{Chord span and deterministic cubic families}
The bound $n(n-1)/(6s)$ isolates the role of chord span, but is quadratic only when $s$ is bounded. Determining the order uniformly for growing $s$ would refine the one-chord example. The largest makespan over bridgeless cubic graphs also remains to be determined: the universal quadratic upper bound and the much smaller bounds for typical cubic graphs do not settle that class maximum. Fixed regular degrees at least four provide further unresolved families.

\paragraph{Connectivity and common neighbors}
For $k$-vertex-connected graphs with $k\ge3$, Section~\ref{app:connectivity} proves an upper bound of
\[
 O\!\left(\min\left\{\frac{n^2}{k},\frac{n^2}{k^2}\frac{\log n}{\log\log n}\right\}\right).
\]
A family of lower bounds gives $\Omega(n/k)$ for $3\le k\le n/2$, leaving a substantial gap. The constant-order bounds under strong common-neighbor hypotheses in Section~\ref{sec:extensions} suggest studying which local intersection conditions suffice for low-cost simultaneous exchanges beyond the dense setting.

\paragraph{Planar graphs and lattices}
The linear class maximum for simple 3-connected planar cubic graphs uses the way face boundaries intersect in these graphs. Both cubicity and 3-connectivity enter the proof. The corresponding class maximum for bridgeless planar cubic graphs is open. A vacancy-only transfer in planar graphs, with a bound expressed through distances and cut capacities, would address a different limitation of the current facial proof. For standard honeycomb patches, an $O(a+b)$ matching-routing scheme or a suitable bounded-congestion simulation of a square grid would match the diameter lower bound. We include the square and triangular lattice estimates in Section~\ref{app:lattice} to recover established benchmarks from prior work rather than to introduce new bounds.

\section*{Acknowledgments}
This work was carried out while one of the authors, Lucy McManamon, was a participant in the 2026 DIMACS REU program at Rutgers University, supported by NSF grant CCF-2447342.

\bibliographystyle{siamplain}
\bibliography{references}
\end{document}